\documentclass[twocolumn]{aastex631}
\usepackage{amsmath}
\usepackage{lmodern}
\hypersetup{pdftitle={When the Shadow Meets Its Measure: Assessing the Feasibility of Submillimeter Black Hole Shadow Imaging in Megamaser-disk Active Galactic Nuclei},pdfauthor={Roman N. Burridge and Geoffrey C. Bower},pdfsubject={Updated author manuscript for The Astrophysical Journal 1011, 273 (2026); doi:10.3847/1538-4357/ae8a42}}
\journalinfo{Updated author manuscript. ApJ 1011, 273 (2026); doi:10.3847/1538-4357/ae8a42}
\graphicspath{{./}}

\begin{document}

\title{When the Shadow Meets Its Measure:\\Assessing the Feasibility of Submillimeter Black Hole Shadow Imaging in Megamaser-disk Active Galactic Nuclei}
\author[0009-0000-3265-7726]{Roman N. Burridge}
\affiliation{Department of Physics and Astronomy, University of Hawai'i at M\=anoa, 2505 Correa Road, Honolulu, HI 96822, USA}
\author[0000-0003-4056-9982]{Geoffrey C. Bower}
\affiliation{East Asian Observatory, 660 N. A'ohoku Place, Hilo, HI 96720, USA}
\affiliation{Institute of Astronomy and Astrophysics, Academia Sinica, 645 N. A'ohoku Place, Hilo, HI 96720, USA}
\affiliation{Department of Physics and Astronomy, University of Hawai'i at M\=anoa, 2505 Correa Road, Honolulu, HI 96822, USA}
\correspondingauthor{Roman N. Burridge}
\email{rburr@hawaii.edu}

\begin{abstract}
Well-modeled water megamaser disks offer a path to black hole shadow (BHS) tests of strong-field gravity beyond Sgr~A* and M87*: their very long baseline interferometry (VLBI) angular dynamics precisely constrains $M_\bullet/D$, the maser-disk orientation, and the dynamical center within the maser reference frame, thereby predicting the fiducial Schwarzschild BHS angular size without an externally adopted distance. We assess 21 megamaser galaxies as BHS targets using published maser measurements to calculate BHS angular sizes and baseline requirements. We combine new Submillimeter Array (SMA) continuum observations with archival Atacama Large Millimeter/submillimeter Array (ALMA) and Very Large Array (VLA) measurements to estimate arcsecond-scale 230~GHz flux densities and identify the strongest nuclear-region sources for higher-resolution follow-up. We also estimate the maximum dust and extended-jet contributions, examine limited variability measurements, and, for the leading candidate NGC~4258, evaluate the astrometry required to detect a spin-dependent offset between the BHS center and the maser dynamical center. Only NGC~4258 has an adopted BHS angular size resolvable with an Earth--L2 baseline; the other adopted sizes require longer baselines, while an Earth--L4/L5 separation would be sufficient for all. Five sources reach $S_{230}\gtrsim10$~mJy. However, whether the observed 230~GHz emission arises on BHS scales and is sufficiently optically thin to reveal a BHS remains unknown throughout the sample. NGC~4258's submillimeter excess is consistent with a thin disk transitioning to an advection-dominated flow; its spin-offset test remains infeasible with present maser astrometry.
\end{abstract}


\section{Introduction}
Measuring black hole (BH) masses is fundamental both to astrophysics and to tests of general relativity. Mass determines the gravitational radius that defines the event horizon, and recent horizon-scale imaging has resolved the black hole shadow (BHS)\footnote{We use ‘BHS’ to refer to the region on the observer’s screen corresponding to photons that are captured by the BH, leaving a dark area against the surrounding emission. It is circular in the Schwarzschild case, and displaced or distorted by spin and viewing inclination in the Kerr case.} in M87* and Sagittarius A*, in agreement with the predictions of general relativity  \citep{EventHorizonTelescopeCollaboration2019f,EventHorizonTelescopeCollaboration2022d,EventHorizonTelescopeCollaboration2022e,EventHorizonTelescopeCollaboration2022f}. Mass also anchors the scaling relations that link supermassive black holes (SMBH) to the growth of their host galaxies, and sets the Eddington limit that regulates accretion and feedback \citep{Kormendy2013,Fabian2012,Heckman2014}. Traditional dynamical methods, including stellar and gas kinematics, reverberation mapping, single-epoch virial methods, and water megamaser disk modeling with very long baseline interferometry (VLBI) maps of the maser distribution, probe the motions of tracers in the gravitational potential to infer a BH mass estimate. By contrast, submm-mm VLBI of the horizon-scale emission delivers a geometric probe of spacetime itself.

Direct imaging of a BHS with submillimeter/millimeter (submm-mm) VLBI requires AGN-core emission surrounding the SMBH to be (1) large enough in angular extent to be resolved, (2) sufficiently bright, and (3) optically thin at the observing frequency. In this work, we evaluate AGN hosting water megamaser disks using two preliminary target-selection metrics: predicted angular BHS size and available arcsecond-scale submm-mm continuum flux density.

We selected 21 water megamaser AGN that are among the best characterized in the Megamaser Cosmology Project (MCP) and related VLBI literature. Published 22~GHz VLBI observations resolve their nuclear maser distributions and provide source-specific dynamical constraints. For 19 sources, adopted maser dynamical results constrain $M_\bullet/D$ and permit calculation of predicted BHS sizes; Mrk~1029 and NGC~1320 remain too complex for robust mass determinations and therefore have no adopted BHS sizes. The full sample, adopted parameters, and primary references are listed in Table~\ref{tab:md}. The expected BHS angular size is \citep{Pesce2021}:
\begin{equation}
\theta_{\rm BHS} = \sqrt{27}\frac{R_S}{D_{\mathrm{SMBH}}}, \quad R_S = \frac{2GM}{c^2}
\label{eq:bhs-angle}
\end{equation}
where $M$ is the BH mass, $D_{\mathrm{SMBH}}$ is its distance from Earth, $R_S$ is the Schwarzschild radius, and the prefactor $\sqrt{27}$ corresponds to a Schwarzschild BH \citep{Bardeen1973}.

Robust maser VLBI positions and rotation speeds constrain $M_\bullet/D$ through Kepler's third law. This ratio sets the black hole's angular gravitational scale, while the same observations constrain the disk orientation and dynamical center within the maser reference frame \citep[e.g.,][]{Kuo2010,Reid2019,Pesce2020a}. Equation~\ref{eq:bhs-angle} converts this angular scale into the fiducial Schwarzschild BHS angular size without an externally adopted distance. Comparing that weak-field disk-dynamics prediction with an observed shadow links maser dynamics directly to strong-field imaging \citep{EventHorizonTelescopeCollaboration2022f}. Neither an adopted distance nor its uncertainty enters this comparison; uncertainty in $M_\bullet/D$ and in the maser and shadow models remains. General-relativistic modeling of the maser kinematics in NGC~4258 and NGC~1194 has also identified potentially measurable gravitational-redshift contributions, providing a complementary relativistic test using the maser spectra themselves \citep{Nucamendi2021,GonzalezJuarez2024}.

The predicted angular sizes translate into demanding interferometric baseline requirements. Previous analyses across the broader AGN population likewise find that resolving additional BHS targets beyond M87* and Sgr~A* generally requires submm-mm baselines beyond current ground-based EHT scales \citep{Zhang2025,Silpa2026}. Values of $\theta_{\rm BHS}$ for the water maser sample are listed in Table~\ref{tab:md}. NGC~4258, the largest at $0.54~\mu$as, requires a baseline roughly 40 Earth diameters and is therefore the leading sample target for space-based VLBI (SVLBI). RadioAstron has already detected its 22~GHz H$_2$O maser emission on Earth--space baselines extending to nearly 20 Earth diameters, resolving the emission near the systemic velocity\footnote{In megamaser disk terminology, ``systemic'' maser features lie near the host galaxy's systemic velocity and arise from gas projected close to the line of sight toward the nucleus. By contrast, ``high-velocity'' features are substantially blueshifted or redshifted relative to systemic velocity and trace the approaching and receding sides of the rotating disk.} into compact components with high brightness at an orbital radius of $\simeq0.126$~pc \citep{Baan2022}.

To assess the VLBI baselines required to resolve the BHS of our set of megamaser disk AGN, we estimate the angular resolution of several Earth to space baselines using the diffraction limit:
\begin{equation}
\theta_{\mathrm{VLBI}} = \frac{\lambda}{D_{\mathrm{BL}}}
\label{eq:diffraction-limit}
\end{equation}
where $\theta_{\mathrm{VLBI}}$ is the interferometric fringe spacing, $\lambda$ is the observing wavelength, and $D_{\mathrm{BL}}$ is the interferometric baseline length. We adopt 230~GHz as the reference frequency when calculating the baseline resolving power. In addition to the Earth-diameter baseline achieved by the Event Horizon Telescope (EHT), we consider several existing or proposed space--ground configurations: the \emph{BHEX} and \emph{Millimetron} submm-mm mission concepts, a proposed VLBI modification of the L2 \emph{Origins} concept, and the X-band \emph{LOVEX} lunar-orbit experiment \citep{Johnson2024,Andrianov2020,Syachina2024,Pesce2019,Hong2025}. \emph{LOVEX} demonstrates a lunar-orbit space--ground baseline approaching the Earth--Moon distance, which we use here only as a geometric benchmark at 230~GHz. \emph{Millimetron} and the proposed interferometric use of \emph{Origins} motivate the Earth--L2 benchmark \citep{Andrianov2020,Pesce2019}. For completeness, the Earth--L4/L5 baseline serves as a fiducial long-baseline geometry capable of resolving all adopted primary BHS sizes in our megamaser disk AGN sample.

At 230~GHz, the corresponding angular resolutions are 21.1~$\mu\mathrm{as}$ for an Earth-diameter baseline (12,756~km), 10.1~$\mu\mathrm{as}$ for the representative BHEX baseline (approximately 26,600~km), 0.69~$\mu\mathrm{as}$ for an Earth--Moon space--ground baseline (approximately 389,000~km), 0.18~$\mu\mathrm{as}$ for an Earth--L2 baseline (1.5~million~km), and 0.0018~$\mu\mathrm{as}$ for an Earth--L4/L5 baseline (150~million~km).

At the Earth--L2 benchmark, only the BHS of NGC~4258 is large enough to be resolved at 230~GHz. The next largest BHSs, those of NGC~1194 and NGC~1068, would require an increase in baseline distance to $\sim$ 1.5 times the distance from Earth to L2. All adopted primary BHS sizes in the water maser AGN sample would be resolved with a baseline extending from Earth to L4/L5, which serves as an illustrative long-baseline configuration. We therefore discuss six sources in detail: NGC~4258, NGC~1194, and NGC~1068 have the largest predicted BHSs, while NGC~3079, NGC~4945, and Circinus provide the strongest submm-mm continuum estimates among the remaining sources.

Having considered the baseline and flux-density requirements for resolvability, it is also essential to consider how the viewing inclination of the emitting accretion disk flow relative to the BH spin axis influences the appearance of the BHS. This viewing inclination couples with spin and optical depth to determine the degree of asymmetry in the observed image, introducing a fundamental degeneracy among these three parameters\footnote{In a moderately optically thick regime, a non-spinning SMBH viewed edge-on appears semicircular, whereas in the optically thin regime, the BHS remains circular regardless of viewing inclination. In both cases, spin introduces asymmetries and a centroid displacement.} \citep{Takahashi2004}. In the fully optically thick regime, the BHS is completely obscured and cannot be recovered. In the moderately optically thin regime, photon rings\footnote{Photon rings are photons that orbit near the critical curve before escaping, forming narrow rings that converge exponentially onto the BHS’s edge.} remain inaccessible, but the BHS becomes marginally resolvable as synchrotron photons from the hot, magnetized accretion flow begin to escape \citep{Inoue2014}. At shorter wavelengths, as the emission becomes increasingly optically thin, the photon ring can be resolved \citep{Bronzwaer2020, KianaSalehi2024}. Each successive subring is progressively demagnified, rotated, and time-delayed relative to the last. Recovering the corresponding observables -- the Lyapunov exponent, time delay, and azimuthal shift -- therefore requires not only sufficient resolution, but also a reliable determination of the line-of-sight viewing geometry of the emitting accretion disk flow. For the 21-source sample, the inclination entries in Table~\ref{tab:md} come from maser observations or interpretations of the maser geometry; the Sgr~A* and M87* comparison values instead come from EHT modeling and the jet axis. Where the maser geometry can be modeled, it provides a direct subparsec inclination estimate. Otherwise, Table~\ref{tab:md} describes the geometry qualitatively or as unconstrained. Quoted errors do not include all geometric and model uncertainty, particularly for sparse or complex systems \citep{Reid2019,Gallimore2023,Pesce2020b,Pesce2020a,Reid2013,Kuo2015}. If the maser disk traces the orientation of the inner accretion flow, its inclination supplies an external geometric reference for interpreting BHS and photon ring morphology; if it does not, the comparison itself reveals a physically important warp or misalignment. In systems with resolved maser substructure, such mismatches can point to warped or magnetically influenced accretion flows that must be modeled before the maser disk is treated as a direct tracer of the innermost accretion flow geometry \citep[e.g.,][]{Gallimore2024}. The 10-100~pc measurements trace circumnuclear gas well beyond that sub-parsec maser disk, so the two orientations need not match each other or the BH spin axis. Comparing the sub-parsec maser disk with the 10-100~pc circumnuclear gas reservoir can therefore test whether the inner disk is aligned with the gas supply on larger scales or instead more closely aligned with the BH spin axis, thereby helping to distinguish whether the system resides in a standard and normal evolution (SANE) or magnetically arrested disk (MAD) magnetic accretion flow state \citep{Ressler2023}.

The viewing inclination of the emitting accretion disk flow is difficult to extract from VLBI images of Sgr~A* because of degeneracies with accretion and outflow properties. For M87*, it must be inferred from the jet axis rather than from the BHS itself \citep{EventHorizonTelescopeCollaboration2021b,EventHorizonTelescopeCollaboration2022f}. Estimates from other tracers span widely different spatial scales -- stellar disks on kiloparsec scales, molecular gas on hundreds of parsecs, and infrared or X-ray obscurers on parsec scales -- and carry large uncertainties \citep{Efstathiou2021, Murphy2011, Brightman2011}.

Stellar dynamics, gas dynamics, and single-epoch virial estimates provide BH masses but generally depend on an adopted distance. Direct reverberation mapping instead sets the broad-line-region scale from a measured time delay and can yield a distance-independent mass, although its accuracy depends on the broad-line-region geometry and inclination \citep{Blandford1982,Pancoast2014}. For a megamaser AGN with a distance-independent reverberation mass and a maser measurement of $M_\bullet/D$, combining the two would yield an independent distance. The cleanest 22~GHz water maser disks can provide a more direct geometric constraint because their compact maser spots trace ordered, nearly Keplerian rotation in molecular disks on sub-parsec scales. The adopted maser-dynamical masses are not all obtained identically: some analyses explicitly fit a central point mass, whereas others infer the total gravitating mass enclosed by the maser orbits and attribute it primarily to the SMBH. Disk self-gravity, noncircular structure, and sparse sampling can therefore shift or broaden the inferred SMBH mass and BHS scale. A flatter-than-Keplerian rotation curve can indicate a massive disk, in which case treating all enclosed mass as the SMBH would overestimate $M_\bullet$; spiral structure, as inferred in NGC~1068, can also produce an apparently flattened curve \citep{Kondratko2005,Gallimore2024}. Among the six sources discussed individually, these limitations include the spiral structure of NGC~1068, the thick self-gravitating disk of NGC~3079, the sparse maser sampling of NGC~4945, and a possible disk contribution in Circinus. Their consequences are addressed in the individual source discussions. The 22~GHz maser emission arises from a collisionally pumped population inversion in warm ($\sim$\,300-1000~K), dense ($n_{\mathrm{H}_2} \sim 10^{7}\text{-}10^{11}\,\mathrm{cm}^{-3}$) gas with sufficient H$_2$O column density and line-of-sight velocity coherence; in disk maser systems, systemic features can amplify background AGN continuum, whereas high-velocity features are generally interpreted as self-amplified emission along paths with coherent velocity in the rotating disk \citep{Lo2005}.

The six sources discussed in detail span a range of dynamical and geometric constraints, allowing us to assess how confidently their BHS sizes can be predicted. Table~\ref{tab:md} reports adopted BHS uncertainties for NGC~4258 and NGC~1194 and central estimates for the other four, whose available uncertainty information is incomplete or model dependent.

This paper is organized as follows. Section~\ref{sec:data} describes the archival compilation of SMBH mass, distance, and maser disk orientation; newly conducted SMA observations; ALMA and VLA archival compilation; and Common Astronomy Software Applications (CASA) fitting for continuum extraction. Section~\ref{sec:results} presents the primary candidates of megamaser disk AGN for BHS imaging using SVLBI; evaluates the proposed spin-offset measurement in NGC~4258 by separating the formal 230~GHz BHS centroid precision from the limiting uncertainty in 22~GHz maser astrometry; and presents the flux densities of the sample at submm-mm wavelengths, including contamination bounds and a variability check on the reliability of the continuum estimates. Section~\ref{sec:individual-sources} discusses these six sources individually. Section~\ref{sec:conclusion} summarizes our conclusions.

\section{Data Collection}
\label{sec:data}
This section describes the mass and distance compilation, continuum observations, archival measurements, and uniform fitting procedure used in the analysis.

\subsection{Mass, Distance, and Maser Disk Orientation}
\label{subsec:md}
The sample comprises the 21 megamaser AGN in Table~\ref{tab:md}. Nineteen have adopted maser dynamical constraints on $M_\bullet/D$, represented in the table by masses and distances, from which Equation~\ref{eq:bhs-angle} gives the predicted BHS diameters used in Figure~\ref{fig:BHSflux}; Mrk~1029 and NGC~1320 have neither robust adopted masses nor BHS sizes. Each reported mass is matched to the distance adopted here. Hubble-flow distances retain the estimator used by the cited mass analysis and are standardized to $H_0=73.9~\mathrm{km\,s^{-1}\,Mpc^{-1}}$ \citep{Pesce2020b}. For distances not derived from the Hubble flow, we adopt published values and uncertainties; the source-specific references are given in Table~\ref{tab:md}.

For thirteen systems, a published maser analysis gives a physical mass at a reference distance $D_0$. Kepler's third law relates the measured orbital velocity and angular disk radius to $M_\bullet/D$. When a different distance $D$ is adopted, the physical mass and its mass-fit uncertainty are rescaled by $D/D_0$. Because $M_\bullet$ and $D$ change by the same factor, $M_\bullet/D$, and therefore the predicted angular BHS size, remains unchanged.

For the other six systems---NGC~4258, CGCG~074--064, UGC~3789, NGC~6264, NGC~6323, and NGC~5765b---published Markov Chain Monte Carlo (MCMC) disk fits allow $M_\bullet$ and $D$ to vary together, providing a more complete statistical description of $M_\bullet/D$. For every paired mass--distance sample, we calculate $\theta_{\rm BHS}$. We adopt the median and central 68\% interval of the resulting $\theta_{\rm BHS}$ distribution. These summaries were calculated from the chains used for the published results \citep{Reid2019,Pesce2020a,Pesce2020b} and supplied by D.~W. Pesce (2026, private communication). Table~\ref{tab:md} reports the central 68\% interval as the statistical uncertainty and separately allows the fitted $M_\bullet/D$, and therefore the BHS size, to be 1\% higher or lower because of small imperfections in the Keplerian disk model.

The mass and distance columns of Table~\ref{tab:md} retain their published uncertainties. Across the sample, Table~\ref{tab:md} adopts a primary BHS uncertainty when supported by a paired posterior or complete published estimate and lists separately quantified systematic or model terms in their own column.

\subsection{SMA Observation and Calibration}
\label{subsec:smaobs}
We conducted submm-mm continuum observations using the Submillimeter Array (SMA) on Maunakea in the subcompact configuration on 2023 March 16 under good weather conditions, with precipitable water vapor (PWV) below 4~mm, under program 2022B-H002. Out of 8 antennas, antenna2 and antenna7 were unavailable. The lower sideband (LSB) ranges from 209.4-221.7~GHz, and the upper sideband (USB) from 229.4-241.7~GHz, providing 24.6~GHz of aggregate frequency coverage across the two sidebands. Each sideband consists of six 2.3~GHz spectral windows. Of the 21 sources in our sample, 15 were observed. J0437+2456, IC~2560, NGC~3393, and CGCG~074-064 were not included in this run, and Circinus and NGC~4945 were outside the observable range. The synthesized beam size was approximately 5-6~arcseconds full width at half maximum (FWHM). Calibration began with 3C84 (bandpass) and Uranus (flux) and concluded with 3C279 (bandpass) and Ceres (flux). Phase calibrators were selected based on flux density ($\geq$~0.5~Jy) and angular separation ($\leq$~15\textdegree) from each science target.

We also conducted a submm-mm observation for NGC~4258 using the SMA on 2022 Jan 31 under satisfactory weather conditions, with PWV$<4$~mm, under program 2021B-H004. Out of 8 antennas, antenna3 and the lower sideband for antenna2 were unavailable. The lower sideband ranges from 219.4-231.7~GHz, and the upper sideband from 239.4-251.7~GHz. The synthesized-beam FWHM was approximately 3~arcseconds. The flux calibrator chosen for this observation was MWC~349a, and the bandpass calibrator was 3C~279. This dataset was not included in the final analysis because the clean image products did not satisfy our adopted image selection criteria.

We calibrated and imaged the raw visibility data in CASA\footnote{SMA reduction is outlined in \url{https://lweb.cfa.harvard.edu/rtdc/SMAdata/process/tutorials/sma_in_casa_tutorial.html}}. In addition to flagging edge channels, we flagged common extragalactic molecular line regions brighter than 1~Jy as listed by \citet{Martin2021}. The selected lines in the two SMA tunings were C$^{18}$O at 219.560~GHz, $^{13}$CO at 220.399~GHz, CN hyperfine components at 226.660, 226.874, and 226.875~GHz, CO at 230.538~GHz, and CS at 244.936~GHz. For each target, the rest frequencies were shifted to the observed frame over the quoted systemic-velocity uncertainty, and the resulting interval was padded by 300~km~s$^{-1}$ on each side. After overlapping intervals were merged, the excluded channels spanned approximately 0.9-2.2~GHz per target, or $\sim$4-9\% of the 24.6~GHz aggregate sideband coverage.

The reference antenna was set to antenna8, which established a zero point for the phase. In CASA \texttt{gaincal}, the short solution interval used \texttt{solint='int'}, producing one solution per integration (14.84~s in 2022B-H002 and 29.68~s for the 2021B-H004 science and gain calibrator scans). The long solution interval used \texttt{solint='inf'}, yielding one solution per scan; the nominal science scans were 445.2~s in 2022B-H002 and 890.5~s in 2021B-H004, while the corresponding phase calibrator scans were generally 59.4--118.7~s and 178.1~s, respectively. For a given frequency channel and antenna, we required the calibration tables to retain only solutions for which at least three baselines successfully produced solutions. We also required a minimum signal-to-noise ratio (SNR) of 2 for phase calibration and 3 for amplitude calibration.

First, we generated a bandpass calibration table from 3C~279 while using self-calibration to correct short-term phase variations. Long- and short-term phase calibration tables were then generated for the phase, flux, and bandpass calibrators while applying the bandpass table. To maximize the SNR of the continuum image, spectral windows were combined in the upper and lower sidebands. Long-term amplitude calibration tables were then generated for the phase, flux, and bandpass calibrators while applying the short-term phase and bandpass correction tables. The flux calibrator was fixed at its known flux density to establish an absolute flux scale across the bandwidth, and flux calibration tables were generated for the phase and bandpass calibrators. The flux calibrator for 2022B-H002 was Uranus and was modeled using the Butler-JPL-Horizons 2012 flux standard. The flux calibrator for 2021B-H004 was MWC~349a and was assigned a manual flux density of 1.98 Jy based on its value in the SMA Calibrator Catalog.

The science target images were first inspected before CLEAN deconvolution. A source was retained as an SMA detection only if the dirty image before CLEAN iterations reached SNR~$\geq 3$ at the source position. Retained detections were imaged with CASA \texttt{tclean} in multifrequency synthesis mode, combining the continuum channels left after edge and molecular line flagging into a single image with 1\arcsec{} pixels. We used the CASA default CLEAN gain of 0.1. The final CLEAN depth was chosen by inspecting the residual images; we adopted the image in which the brightest source region residual was comparable to the measured image root-mean-square (RMS) noise.

\subsection{Archival Continuum Measurements}
\label{subsec:archcoll}
To supplement our SMA observations, we retrieved public continuum images from the Atacama Large Millimeter/submillimeter Array (ALMA) and Karl G. Jansky Very Large Array (VLA) archives using \texttt{astroquery}. The released continuum table (Table~\ref{tab:mr1}) contains 501 measurements: 485 image-based rows and 16 literature measurements. For each image, we obtained the observing frequency and synthesized-beam dimensions from its header with the CASA task \texttt{imhead}. Frequency averaging in an off-axis VLA image can spread the source signal and lower its measured peak. We excluded nine measurements potentially affected by this bias. The quantitative selection criterion, archive identifiers, and product-recovery information are provided with the released machine-readable table.

Three artifact-dominated NGC~4258 NVAS images were excluded. The machine-readable table identifies the published measurements supplementing the fitted archive products and gives their references \citep{Doi2005,Doi2013,Kamali2017}.

\subsection{CASA Fitting}
\label{subsec:casa}

The continuum extraction used four stages: (1) a preliminary unconstrained Gaussian fit, (2) a free-centroid fit with the Gaussian size and orientation fixed to those of the restoring beam in a broad region centered on the NED position, (3) selection of an eligible high-resolution 200--400~GHz detection, whose beam-fixed Gaussian fit was designated the anchor fit for that source, and (4) repetition of the beam-fixed fit for every retained image within a smaller region centered on the anchor-fit centroid. The anchor image was an SNR~$\geq3$ detection with beam FWHM $<1\arcsec$ and, when several qualified, the image with the smallest beam. The centroid of the anchor fit defined the anchor position. When no anchor image was available, the final search remained centered on the NED position.

For the final fits, we fit a Gaussian whose axes and position angle were fixed to the restoring beam and use the fitted component flux density. With perfect relative astrometric registration, the centroid would also have been fixed at the anchor position. To accommodate residual astrometric offsets between images, we instead allowed the centroid to vary, but only within a restricted region centered on the anchor position that was substantially smaller than the broad NED-centered region used in stage (2). NGC~4945 was the sole positional exception: its images were refitted at the maser/AGN position of source~18 from \citet{Emig2020}, rather than the nearby source~17.

After excluding the fitted source, we convert the remaining pixel brightness values to mJy~beam$^{-1}$ and estimate the local image noise as 1.4826 times their median absolute deviation. Fits with SNR~$\geq3$ are detections; otherwise, we report a $3\sigma$ upper limit using this local image-noise estimate. Predesignated dirty SMA images instead retain local RMS limits. The quoted image uncertainties reflect this local noise estimate and do not include absolute calibration, source-model, aperture-choice, or variability terms. Invalid or artifact-dominated fits are excluded. Full fitting parameters and diagnostics are retained with the released workflow.

For the common-frequency comparison, we select the smallest-beam detection from 200--400~GHz for each source, or the smallest-beam upper limit when no detection is available, as the $S_{230}$ reference measurement. We calculate $S_{230}$ for each selected measurement using $S_\nu\propto\nu^\alpha$. We adopt $\alpha=-0.7$, a representative slope for optically thin synchrotron emission, and report values for $-1.0\leq\alpha\leq-0.4$ to show the dependence on the assumed slope \citep{Rybicki1979,Marsden2014}. The fiducial values are plotted in Figure~\ref{fig:BHSflux}. Tables~\ref{tab:NGC4258_flux}--\ref{tab:circinus_flux} give the numerical values and spectral-index ranges for the six primary candidates. Numerical values and ranges for the full 19-source set are provided in the released machine-readable $S_{230}$ reference table. The resulting set contains 12 detections and seven upper limits.

\section{Results and Analysis}
\label{sec:results}
This section combines the predicted BHS sizes and continuum estimates to assess the leading SVLBI targets and the NGC~4258 spin-offset experiment.

\subsection{SMA results}
\label{subsec:smares}
For NGC~4258, we use our SMA detection to estimate $S_{230}$. Our SMA observations also detected NGC~1068 and NGC~3079, but higher-resolution ALMA and NOEMA measurements, respectively, provide the inputs for their $S_{230}$ estimates.

SMA non-detections used to set $S_{230}$ upper limits include NGC~2273, NGC~6264, UGC~6093, NGC~6323, UGC~3789, and Mrk~1029.

SMA non-detections superseded by higher-resolution $S_{230}$ reference measurements include NGC~1194, NGC~2960, NGC~4388, NGC~5765b, ESO~558-G009, and NGC~1320. NGC~4388 was not retained as an SMA detection after the molecular line channels were flagged. Warm molecular gas with noncircular motions is present on roughly 100-pc scales \citep{Greene2014}, so line contamination is a plausible explanation for the pre-flagging SMA feature.

Neither CGCG~074--064 nor J0437+2456 has an eligible continuum measurement from which to estimate $S_{230}$.

The final SMA photometry used beam-shaped Gaussian fits and therefore did not independently measure source size. We treat these detections as unresolved at the SMA resolution for the purposes of this work.

\subsection{Primary Candidates for SMBH Imaging using VLBI}
\label{subsec:primecan}
We assess the targets using their predicted angular sizes and arcsecond-scale $S_{230}$ estimates, comparing these metrics with the angular-resolution benchmarks from Equation~\ref{eq:diffraction-limit} and the baseline-sensitivity estimate from Equation~\ref{eq:baseline-sensitivity}.

VLBI sensitivity can be estimated using Equation~\ref{eq:baseline-sensitivity}, which defines the theoretical thermal noise, \( \Delta S_{ij} \), for a given baseline \citep{Walker1995}:
\begin{equation}
\Delta S_{ij} = \frac{1}{n_s}\sqrt{\frac{SEFD_i SEFD_j}{2 \Delta t \Delta \nu}} \quad \text{Jy}
\label{eq:baseline-sensitivity}
\end{equation}
where $\mathrm{SEFD}_{i,j}$ are the system equivalent flux densities (Jy) of the two antennas forming the baseline; $n_s$ is the system efficiency factor; $\Delta t$ is the coherent integration time (s); and $\Delta \nu$ is the observing bandwidth (Hz).

The SEFD for \emph{Millimetron} is estimated to be 4000~Jy, while the representative ground-station SEFD adopted for this benchmark is 74~Jy for phased ALMA \citep{Likhachev2022, EventHorizonTelescopeCollaboration2019b}. \emph{Millimetron} is expected to operate with a bandwidth of \( \Delta \nu = 4 \)~GHz, and the integration time is limited by the coherence timescale of ground-based telescopes due to atmospheric fluctuations, estimated as \( \Delta t = 10 \)~s \citep{Syachina2024}. Under these parameters, the thermal noise is \( 1.9 / n_s \)~mJy. For 2-bit quantization (\( n_s = 0.88 \)), this corresponds to an SNR of 4.6 for a 10~mJy source \citep{Novikov2021}. Because this benchmark uses phased ALMA, it does not apply to NGC~4258, which lies outside ALMA's observable sky and would require a sufficiently sensitive northern ground array.

The coherence time on a space to ground baseline is jointly constrained by the stability of the onboard clock at the space station and by atmospheric phase variations at the ground station. Recent proposals suggest that source-frequency phase referencing (SFPR) systems, employing simultaneous multifrequency observations (e.g., 86/230/345~GHz), could mitigate atmospheric fluctuations on the ground, potentially extending coherence times from 10~s to nearly an hour and reducing the theoretical thermal noise to about 0.11~mJy \citep{Jiang2022}. If atmospheric coherence on the ground is improved through SFPR, the bottleneck then shifts to the onboard clock; in this regime, any further gains in coherence time would require a spaceborne hydrogen maser.

When SFPR is not used, ground-based atmospheric stability sets the coherence time, and ultrastable quartz oscillator (USO) SVLBI clocks are not expected to impose a shorter limit. At 230~GHz, integration times of up to 10~s are theoretically achievable for SVLBI using USO clocks, avoiding the need for heavier and more costly maser clocks \citep{Burt2025}.

\begin{figure*}
\definecolor{mygreen}{RGB}{34,139,34}
\includegraphics[width=\linewidth]{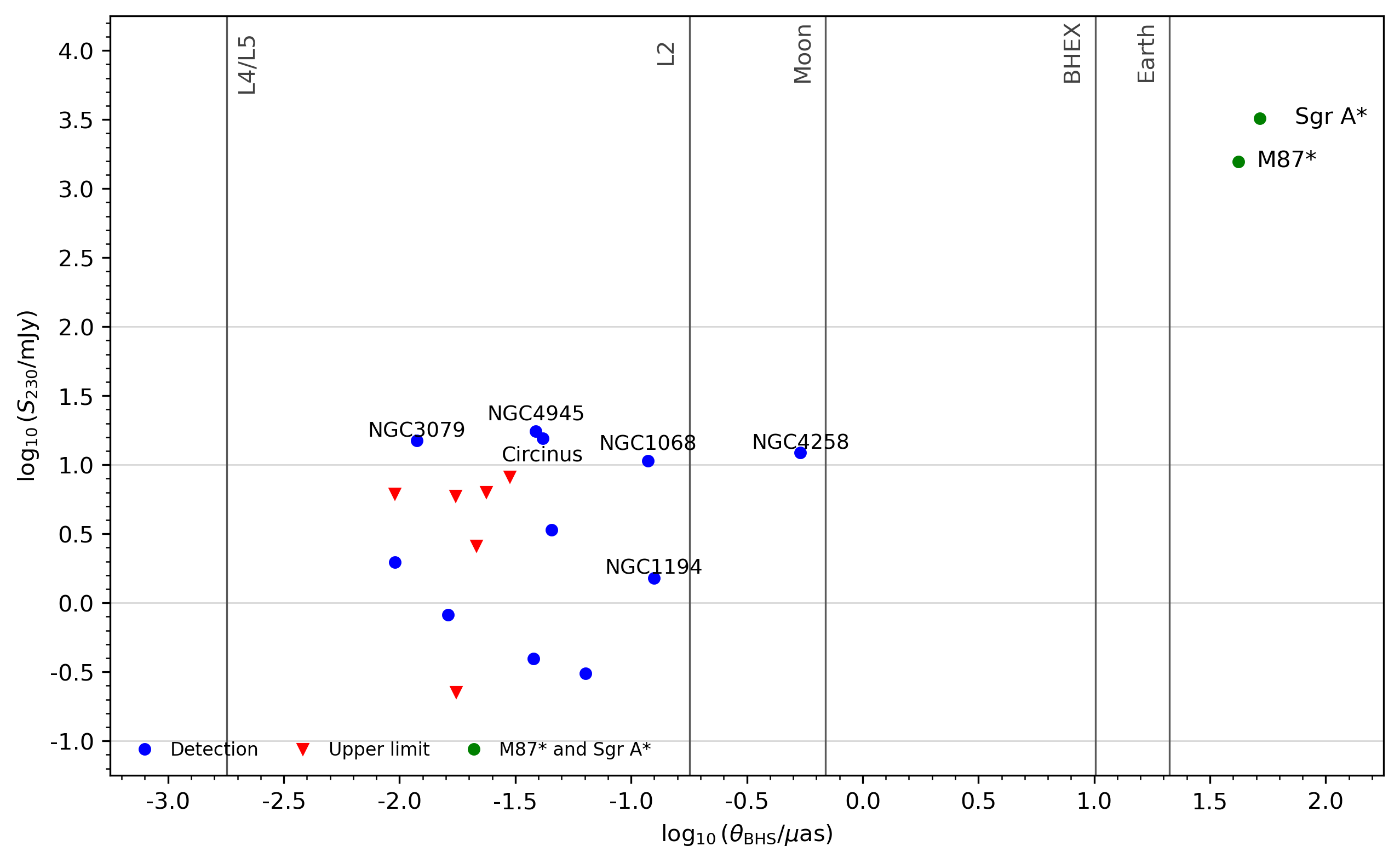}
\caption{Predicted BHS angular size versus nuclear-region $S_{230}$. Blue circles show detections, red triangles show upper limits, and green circles show M87* and Sgr~A* using their long-term mean 230~GHz flux densities. Flux-density uncertainties for the six primary candidates are reported in Tables~\ref{tab:NGC4258_flux}--\ref{tab:circinus_flux}; values for the full sample are provided in the released machine-readable $S_{230}$ reference table. The $\theta_{\rm BHS}$ uncertainties and separate systematic/model terms are reported in Table~\ref{tab:md}. Error bars are omitted here for clarity. Vertical lines mark the nominal 230~GHz resolution of the labeled baselines. Horizontal lines mark representative $S_{230}$ levels across the sample.}
\label{fig:BHSflux}
\end{figure*}

Figure~\ref{fig:BHSflux} compares the targets using nuclear-region $S_{230}$ estimates derived from measurements on angular scales much larger than their predicted BHSs. M87* and Sgr~A* are included for comparison. The BHS values follow Section~\ref{subsec:md} and Table~\ref{tab:md}; the comparison-source flux densities are from \citet{Chen2023} and \citet{Wielgus2022}. VLBI detectability will be set by the correlated flux density on a space--ground baseline, which may be smaller than the plotted $S_{230}$ when nuclear emission is resolved.

NGC~4258 has the largest predicted angular BHS in the megamaser sample. At 230~GHz, an Earth--L2 baseline provides a fringe spacing of $\sim$0.18~$\mu$as, sufficient in angular resolution for its predicted BHS diameter of $0.53860~\mu$as (Table~\ref{tab:md}). Even an Earth--Moon baseline approaches the required resolution. NGC~4258 is therefore the only source in our sample whose predicted diameter exceeds the Earth--L2 fringe spacing. Its arcsecond-scale continuum estimate is $S_{230}\approx12$~mJy (Table~\ref{tab:NGC4258_flux}).

All other sample sources with adopted primary BHS sizes require baselines longer than Earth--L2, with Earth--L4/L5 separations serving as an illustrative upper bound sufficient to resolve all adopted primary BHS sizes in the sample. Among these, NGC~1194 and NGC~1068 exhibit the largest predicted BHS. The five detailed sources with $S_{230}>10$~mJy are NGC~4258, NGC~1068, NGC~3079, NGC~4945, and Circinus.

\subsection{Constraining the SMBH Spin Parameter using SVLBI}
\label{subsec:spin}

Where dual-band receivers allow SFPR, a millimeter/submillimeter image could be directly tied to the 22~GHz water masers, registering the BHS image to the dynamical center inferred from the water-maser kinematics and enabling a spin constraint from their relative offset \citep{Takahashi2004,Bronzwaer2020}. If SFPR is not available, observing both bands simultaneously and referencing them to the same calibrators would still require cross-band astrometric registration. That registration may be affected by additional systematics, including frequency-dependent core shift in a calibrator common to both bands (see Section~\ref{sec:future}). Thus, the spin-offset experiment separates into formal 230~GHz centroiding and relative astrometry; the astrometric component requires both locating the 22~GHz maser dynamical center and registering that point to the 230~GHz BHS image.

This alignment would be simpler if sufficiently compact submm-mm water maser emission could be mapped with VLBI and registered to continuum in the same observing band. Such same-band VLBI registration would avoid the systematic errors introduced by cross-band calibration between 22 and 230~GHz, but it would not remove the requirement for sub-$\mu$as precision in the water maser dynamical center. In evolved stars, including AGB stars and VY~CMa, the 658~GHz water maser has a photon luminosity comparable to or greater than that of the 22~GHz transition \citep{Menten1995,Nesterenok2015}. Models associate the 658~GHz transition with hot, dense molecular gas; its strength in AGN disks will depend on the local temperature, density, radiation field, and water abundance \citep{Gray2015,Nesterenok2015}. 183~GHz lines have been detected in NGC~1194, NGC~3079, NGC~4945, and Circinus, but not in NGC~1068 \citep{Pesce2023, Humphreys2005, Humphreys2016}. 321~GHz lines have been detected in NGC~4945 and Circinus, but not in NGC~1068 \citep{Hagiwara2013, Hagiwara2016, Hagiwara2021, Pesce2016}. Tentative 439~GHz emission has also been reported in NGC~3079 \citep{Humphreys2005}.

The maximal offset between the BHS\footnote{By the black hole shadow center, we mean the midpoint between the shadow's leftmost and rightmost edges.} and the water maser dynamical center is $\simeq 0.725\,R_S$ for an extreme Kerr BH viewed edge-on. Equivalently, this corresponds to $\approx 0.14$ times the BHS diameter \citep{Takahashi2004,Bronzwaer2020}. The angular size of this maximal offset can be predicted directly from the maser $M_\bullet/D$ measurement. Substituting the expression for $R_S$ from Equation~\ref{eq:bhs-angle} and applying the small angle approximation to express the angular shift on the plane of the sky yields $\theta_{\text{shift}} = 0.725 {\theta_{\rm BHS}}/{\sqrt{27}}$.

For NGC~4258, the largest source in the sample, the maximal expected offset between the BHS and the megamaser disk dynamical center is only about $0.075~\mu\mathrm{as}$. Detecting an offset of this size at the $1\sigma$ level therefore requires astrometric precision of order $0.075~\mu\mathrm{as}$. A robust $3\sigma$ detection would require a positional uncertainty smaller by roughly a factor of three. The baseline, observing frequency, and SNR necessary to achieve a given precision can be determined using the positional uncertainty relation given in Equation~1 of \citet{Reid1988}:
\begin{equation}
  \sigma_{\theta} = \left( \frac{4}{\pi} \right)^{1/4} \frac{\theta_{\rm VLBI}}{\sqrt{8 \ln 2}} \frac{1}{\text{SNR}}
\label{eq:pos-uncertainty}
\end{equation}
Here, $\theta_{\rm VLBI}$ is the fringe spacing given in Equation~\ref{eq:diffraction-limit}. Applying Equation~\ref{eq:pos-uncertainty} to an Earth--Moon configuration at 230~GHz, with $\theta_{\rm VLBI} \approx 0.69~\mu$as, shows that an SNR of $\sim 4.1$ would be required for the BHS centroid uncertainty to match the expected $\sim 0.075~\mu\mathrm{as}$ offset; a $3\sigma$ detection would require an SNR of $\sim 12$. These SNR values indicate that formal centroiding of the 230~GHz BHS image is not the dominant limitation. The more stringent requirement is the relative astrometric tie: the 22~GHz water maser dynamical center must be located with sub-$\mu$as precision and registered to the 230~GHz BHS image.

Current 22~GHz water maser VLBI measurements fall well short of this requirement. \citet{Reid2019} determined the dynamical center's x-position as $-0.152 \pm 0.003$ mas and y-position as $0.556 \pm 0.004$ mas. The displacement from the coordinate origin is $\sqrt{x^2+y^2}=0.576$~mas. Propagating the independent coordinate errors along this measured radial direction gives $0.0039$~mas. We round this value to $4~\mu$as and use it as a single astrometric uncertainty scale. This scale is approximately 53 times the maximal $0.075~\mu$as spin-dependent offset. A robust detection would be still more demanding.

The $4~\mu$as scale belongs to the maser dynamical-center solution and is unaffected by a longer 230~GHz baseline. Maser-spot variability, disk turbulence, amplification geometry, and cross-band registration can add further uncertainty. RadioAstron makes the modeling challenge explicit: the systemic 22~GHz emission decomposes into compact, regularly spaced, drifting, and time-variable components on Earth--space baselines \citep{Baan2022}. Progress therefore requires a dynamical-center model that incorporates this structure and its variability.

\subsection{Estimating contamination from thermal dust and extended jets}
\label{subsec:contam}

We estimate potential contamination at 230~GHz by extrapolating measured flux densities using a spectral index $\alpha$ defined by $S_\nu \propto \nu^{\alpha}$, where $S_\nu$ is the flux density at frequency $\nu$ and $\alpha$ is chosen for the relevant emission model. We use only observations with $\nu>1.1\times230$~GHz or $\nu<0.9\times230$~GHz, since such offsets accentuate the spectral slope and enable a clearer separation between rising thermal dust emission and falling extended jet emission. Smaller frequency differences are not used for the contamination extrapolations, but are retained only to assess whether variability could affect the continuum estimates. In Section~\ref{sec:individual-sources}, we present model-dependent estimates of the maximum thermal-dust and extended-jet contributions at 230~GHz.

To estimate contamination from high-frequency thermal dust, we model it as a modified blackbody, $S_\nu=A{\nu^{\beta+3}}/({e^{{h\nu}/({kT})}-1})$ \citep{Casey2012}. Here, $T$ is the dust temperature, $\beta$ is the dust emissivity index, which is set to 1.5, and $A$ sets the model's overall flux-density scale. To test the dependence on dust temperature over a broad range, we evaluate modified-blackbody models from 20 to 2000~K. For each temperature, $A$ is chosen separately so that the model matches the observed higher-frequency flux density before being extrapolated to 230~GHz. The tabulated values retain the archived workflow's power-law projection, $\alpha=\beta+3=+4.5$, for the adopted $\beta=1.5$ \citep{Casey2012}. The modified-blackbody checks do not change any contamination conclusion; the largest difference, for NGC~4258, is described in its individual-source discussion. At the selected beam sizes used for the higher-frequency thermal dust limits (see the beam sizes in the individual source contamination tables and the angular-to-linear scale insets of the SED plots in Section~\ref{sec:individual-sources}), the thermal dust contribution can include both cold dust on galactic scales and warmer circumnuclear dust: cold molecular material can extend over galactic scales, while warmer molecular dust can remain extended on circumnuclear scales. We therefore treat the dust upper limit component as extended across the physical scales sampled by the higher-frequency measurements.

For thermal dust extrapolations, we use only measurements at frequencies $\nu>1.1\times230$~GHz. Each higher-frequency measurement is projected downwards to 230~GHz, and because the dust upper limit component is treated as extended, we additionally apply beam size corrections. If only one thermal dust emission component were present, all projected values would coincide. With this interpretation, any mismatch indicates excess emission not associated with dust. We therefore report the smallest projected value as the modeled maximum thermal-dust contribution under the adopted prescription; its uncertainty is propagated from the selected measurement.

Complementary to the high-frequency thermal dust estimate, we assess low-frequency extended jet emission\footnote{Extended jet refers to emission originating beyond the compact, self-absorbed jet base associated with the AGN and strong-gravity regime.}. We use measurements at frequencies $\nu<0.9\times230$~GHz and project each upward to 230~GHz, assuming a representative optically thin synchrotron slope, $\alpha=-0.5$ \citep{Blandford1978,Rybicki1979}. Treating the extended jet as smaller than the smallest beam size -- consistent with VLBI studies of the water maser AGN sample -- we apply no beam size correction. We therefore report the smallest projected value as the modeled maximum extended-jet contribution under the adopted spectral slope; its uncertainty is propagated from the selected measurement.

Where low-frequency VLBI data are available, we apply the same upward projection to 230~GHz with $\alpha=-0.5$. For literature non-detections, we use the upper limit reported by the cited source or three times its reported $1\sigma$ sensitivity. Where a VLBI image resolves multiple central radio components, we sum their flux densities before projecting them to 230~GHz. If the projection based on VLBI data is smaller than the low-frequency extended jet estimate from archival data, it is adopted. The VLBI extrapolations are therefore not a complete literature survey of compact radio constraints. Additional continuum VLBI exists for NGC~1068 \citep{Gallimore2004, Fischer2023}, NGC~3079 \citep{Irwin1988, Sawada2000, Middelberg2007}, and NGC~4945 \citep{Lenc2009}; high-resolution non-VLBI radio studies also exist for NGC~1068 \citep{Muxlow1996, Mutie2024, Mutie2025}. Maser VLBI/SVLBI detections and searches in the broader source set include NGC~4258 \citep{Baan2022}, NGC~1068 \citep{Greenhill1996, Morishima2022, Gallimore2023}, NGC~3079 \citep{Haschick1990,Trotter1998,Sawada2000,Yamauchi2004,Kondratko2005,Baan2017,Sobolev2018}, Circinus \citep{Greenhill2003,McCallum2009,Sobolev2018}, and NGC~4945 \citep{Greenhill1997}; these constrain maser structure and astrometric registration rather than serving as inputs to the continuum-contamination estimate. Separate VLBI imaging of OH absorption probes the nuclear gas in NGC~3079 \citep{Hagiwara2004}.

\subsection{Variability Check on Continuum Estimates}
\label{subsec:var}

We developed a preliminary pipeline to check whether variability could affect the continuum estimates. Because the archival observations differ in epoch, frequency, and angular resolution, intrinsic variability cannot always be separated from differences in the spatial scales sampled. This check does not change any adopted $S_{230}$ estimate or conclusion presented here; the detailed diagnostic results are provided with the released code \citep{Burridge2026}.

\begin{longrotatetable}
\setlength{\tabcolsep}{2pt}
\begin{deluxetable*}{llllll}
\tablecaption{Predicted Schwarzschild Shadow Diameter, Black Hole Mass, Distance, and Maser Disk Geometry}
\label{tab:md}
\tabletypesize{\footnotesize}
\tablehead{
\colhead{Source} & \colhead{$\theta_{\rm BHS}$ [$\mu\mathrm{as}$]} & \colhead{\shortstack{Systematic/model term\\{}[$\mu\mathrm{as}$]}} & \colhead{$\log_{10}(M_\bullet/M_\odot)$} & \colhead{$D$ [Mpc]} & \colhead{$i$ [deg] / geometry}
}
\startdata
Sgr~A* & $^{1}$\,$51.8 \pm 2.3$ & --- & $^{2}$\,$6.6332 \pm 0.0012$ & $^{1}$\,$0.00815 \pm 0.00015$ & $^{3}$\,$\leq 30$ \\
M87* & $^{4}$\,$42 \pm 3^{\dagger}$ & --- & $^{5}$\,$9.74 \pm 0.03$ & $^{4}$\,$16.8 \pm 0.8$ & $^{6}$\,$\sim 17$ \\
\textbf{NGC~4258} & $0.53860\pm 0.00059$ & $\pm 0.0054$\tablenotemark{a} & $^{7}$\,$7.5999\pm 0.0044$ & $^{7}$\,$7.58\pm 0.11$ & $^{7}$\,$87.05\pm0.09$ (at $r=6.1\,\mathrm{mas}$) \\
\textbf{NGC~1194} & $0.1253\pm 0.0077^{\dagger}$ & --- & $^{8}$\,$7.808\pm 0.043$ & $^{8}$\,$52.6\pm 4.0$ & $^{8}$\,nearly edge-on \\
\textbf{NGC~1068} & $0.11825$ & --- & $^{9}$\,$7.092\pm 0.022$ & $^{10}$\,$10.72\pm 0.52$ & $^{9}$\,$\sim90$ \\
NGC~3393 & $0.0636\pm 0.0041$ & $\pm 0.0091$\tablenotemark{b} & $^{11}$\,$7.498\pm 0.055$ & $^{11}$\,$50.7\pm 5.5$ & $^{11}$\,nearly edge-on \\
NGC~4388 & $0.0454$ & --- & $^{8}$\,$6.924\pm 0.049$ & $^{8}$\,$19.0\pm 2.1$ & $^{8}$\,unconstrained \\
\textbf{Circinus} & $0.04152$ & $+0.0012$\tablenotemark{b} & $^{12}$\,$6.230\pm 0.078$ & $^{12}$\,$4.20\pm 0.80$ & $^{12}$\,warped disk + outflow \\
\textbf{NGC~4945} & $0.0388$ & $+0.0093$\tablenotemark{b} & $^{13}$\,$6.15$ & $^{13}$\,$3.70$ & $^{13}$\,nearly edge-on \\
NGC~5765b & $0.03792\pm 0.00013$ & $\pm 0.00038$\tablenotemark{a} & $^{14}$\,$7.618\pm 0.020$ & $^{14}$\,$112.2\pm 5.3$ & $^{14}$\,$72.4\pm0.5$ (at center) \\
NGC~2273 & $0.0299\pm 0.0020^{\dagger}$ & --- & $^{8}$\,$6.870\pm 0.068$ & $^{8}$\,$25.4\pm 3.5$ & $^{8}$\,nearly edge-on \\
CGCG~074--064 & $0.02835\pm 0.00019$ & $\pm 0.00028$\tablenotemark{a} & $^{15}$\,$7.384\pm 0.039$ & $^{15}$\,$87.6\pm 7.7$ & $^{15}$\,$90.8\pm1.3$ (at center) \\
UGC~3789 & $0.023706\pm 0.000081$ & $\pm 0.00024$\tablenotemark{a} & $^{14}$\,$7.076\pm 0.035$ & $^{14}$\,$51.5\pm 4.3$ & $^{14}$\,$84.9\pm0.6$ (at center) \\
NGC~6264 & $0.021456\pm 0.000055$ & $\pm 0.00021$\tablenotemark{a} & $^{14}$\,$7.441\pm 0.064$ & $^{14}$\,$132\pm 19$ & $^{14}$\,$91.3\pm2.3$ (at center) \\
NGC~2960 & $0.017575$ & --- & $^{16}$\,$7.087\pm 0.027$ & $^{8}$\,$71.3\pm 4.5$ & $^{8}$\,nearly edge-on \\
UGC~6093 & $0.01745\pm 0.00068$ & $\pm 0.00087$\tablenotemark{c} & $^{17}$\,$7.388\pm 0.034$ & $^{17}$\,$143.7\pm 6.7$ & $^{17}$\,$93.8\pm4.0$ \\
ESO~558--G009 & $0.01621\pm 0.00037^{\dagger}$ & $\pm 0.00097$\tablenotemark{d} & $^{18}$\,$7.225\pm 0.036$ & $^{18}$\,$106.3\pm 5.5$ & $^{18}$\,$98\pm1$ (geometry estimate) \\
\textbf{NGC~3079} & $0.0119$ & $+0.0040$\tablenotemark{b} & $^{19}$\,$6.30$ & $^{19}$\,$17.3\pm 1.5$ & $^{19}$\,thick, self-gravitating disk \\
NGC~6323 & $0.009549\pm 0.000059$ & $\pm 0.000095$\tablenotemark{a} & $^{14}$\,$7.01\pm 0.12$ & $^{14}$\,$109\pm 29$ & $^{14}$\,$91.5\pm0.3$ (at center) \\
IC~2560 & $0.00954$ & --- & $^{20}$\,$6.617\pm 0.063$ & $^{20}$\,$44.5\pm 6.1$ & $^{21}$\,nearly edge-on \\
J0437+2456 & $0.00419$ & --- & $^{22}$\,$6.421\pm 0.042$ & $^{18}$\,$64.5\pm 4.3$ & $^{22}$\,unconstrained \\
Mrk~1029 & \nodata & --- & \nodata & $^{18}$\,$119.3\pm 5.9$ & $^{18}$\,$79\pm2$ (geometry estimate) \\
NGC~1320 & \nodata & --- & \nodata & $^{18}$\,$33.8\pm 3.7$ & $^{18}$\,unconstrained \\
\enddata
\end{deluxetable*}
\end{longrotatetable}
\tablecomments{Boldface identifies the six sources discussed individually in Section~\ref{sec:individual-sources}.  Unmarked uncertainties accompanying $\theta_{\rm BHS}$ are statistical.  A dagger marks an uncertainty that includes other effects or was published without separating them.  Lettered entries in the systematic/model column denote (a) a separate allowance for the fitted $M_\bullet/D$, and therefore the BHS size, to be 1\% higher or lower because of small imperfections in the Keplerian disk model, (b) a model sensitivity, (c) a position systematic, and (d) a geometry systematic; these terms remain separate from the uncertainty accompanying $\theta_{\rm BHS}$.  A leading $+$ denotes a one-sided increase in the predicted BHS size.  Asymmetric intervals are displayed using the average of their upper and lower half-widths; exact bounds and incomplete components remain machine-readable.}
\tablerefs{\footnotesize{(1)~\citet{EventHorizonTelescopeCollaboration2022d}; (2)~\citet{GRAVITYCollaboration2022}; (3)~\citet{EventHorizonTelescopeCollaboration2022f}; (4)~\citet{EventHorizonTelescopeCollaboration2019f}; (5)~\citet{Simon2024}; (6)~\citet{EventHorizonTelescopeCollaboration2021b}; (7)~\citet{Reid2019}; (8)~\citet{Kuo2010}; (9)~\citet{Gallimore2024}; (10)~\citet{Markham2026}; (11)~\citet{Kondratko2008}; (12)~\citet{Greenhill2003}; (13)~\citet{Greenhill1997}; (14)~\citet{Pesce2020b}; (15)~\citet{Pesce2020a}; (16)~\citet{Kuo2018Disk}; (17)~\citet{Zhao2018}; (18)~\citet{Gao2016b}; (19)~\citet{Kondratko2005}; (20)~\citet{Nakai2024}; (21)~\citet{Yamauchi2012}; (22)~\citet{Pesce2021J0437}.}}
\tabletypesize{\scriptsize}
\begin{deluxetable*}{lllllll}{
\tablecaption {Continuum Flux-density Measurements}
\label{tab:mr1}
\tablehead{
\colhead{(1)} & \colhead{(2)} & \colhead{(3)}  & \colhead{(4)} & \colhead{(5)} & \colhead{(6)}  & \colhead{(7)} }
}
\startdata
Circinus & ALMA & 39 & 67.8 & 2024/11/24\_1 & 7.52 & 593 \\
Circinus & ALMA & 46 & 48.2 & 2025/01/04\_1 & 3.26 & 116 \\
Circinus & ALMA & 86 & 22.8 & 2016/07/02\_1 & 1.38 & 98 \\
Circinus & ALMA & 93 & 25.7 & 2021/08/01\_1 & 0.219 & 290 \\
Circinus & ALMA & 93 & 42.3 & 2016/01/14\_1 & 2.64 & 143 \\
Circinus & ALMA & 98 & 35.8 & 2024/11/27\_1 & 2.94 & 116 \\
Circinus & ALMA & 107 & 36.0 & 2025/01/26\_1 & 1.35 & 523 \\
Circinus & ALMA & 108 & 40.6 & 2014/12/27\_1 & 2.40 & 158 \\
Circinus & ALMA & 108 & 62.4 & 2014/07/02\_1 & 12.9 & 54 \\
Circinus & ALMA & 108 & 39.8 & 2016/03/18\_1 & 1.54 & 124 \\
\enddata
\tablecomments{
(1): Source Name
(2): Telescope
(3): Frequency [GHz]
(4): Flux Density [mJy]
(5): Date [yyyy/mm/dd\_x]
(6): Beam Size FWHM [arcsec]
(7): SNR
\\
Ten representative rows are shown; the full 501-row table is provided as \texttt{tab2.txt}. It comprises 485 image-based measurements and 16 literature measurements. Image rows use the beam-shaped fits described in Section~\ref{subsec:casa}; literature rows preserve the reported quantities. An asterisk marks an upper limit, and a date suffix distinguishes additional products from the same day. Full numerical precision remains in the machine-readable table.}
\end{deluxetable*}

\section{Individual Source Discussion}
\label{sec:individual-sources}
\begin{figure*}[t]
\begin{center}
\includegraphics[width=0.52\linewidth]{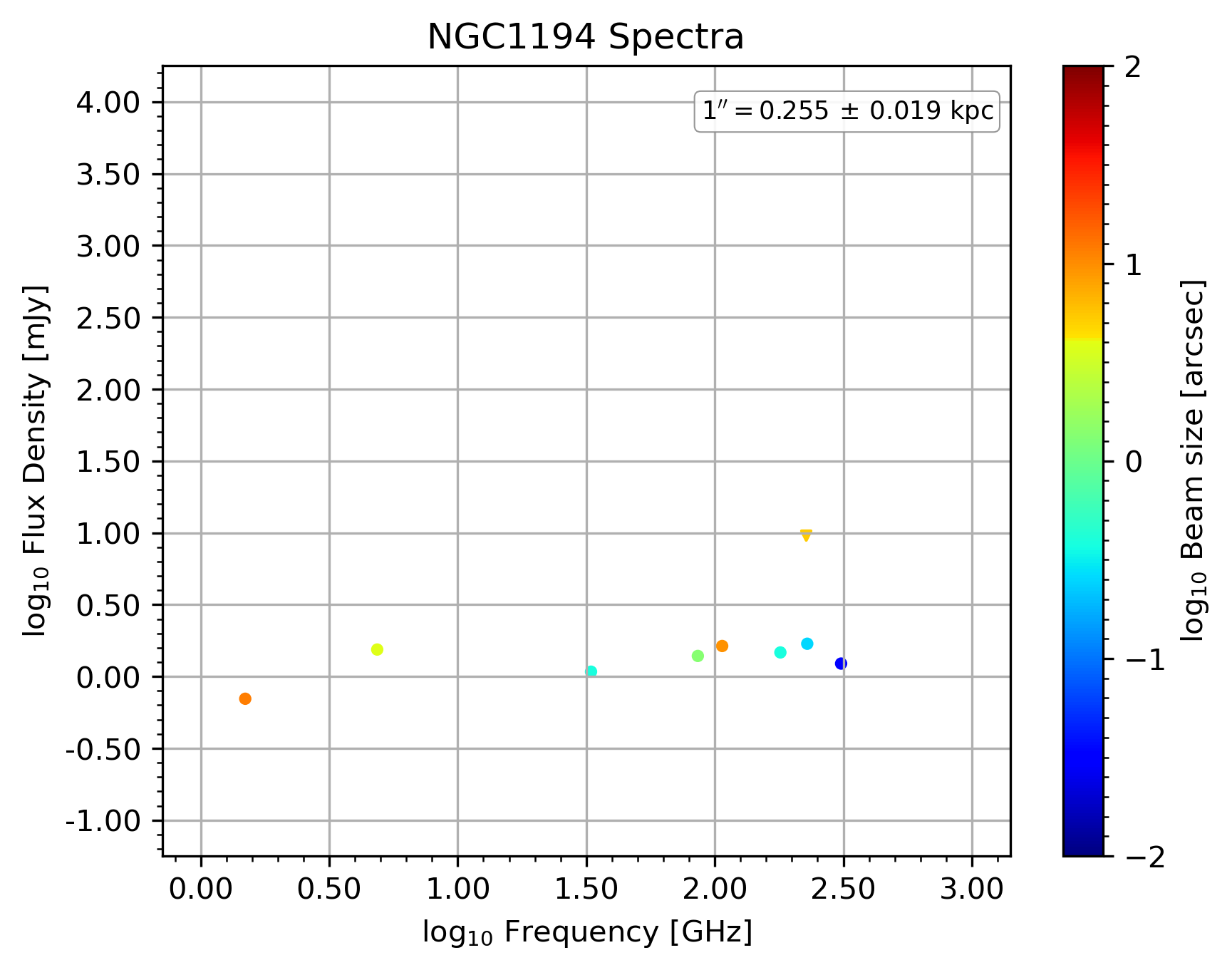}
\caption{Continuum SED of NGC~1194. Circles denote detections and triangles upper limits; color gives the geometric-mean beam FWHM. The inset converts angular to projected physical scale using the adopted distance in Table~\ref{tab:md}.}
\label{fig:sed_1194}
\end{center}
\end{figure*}

\begin{figure*}[t]
\begin{center}
\includegraphics[width=0.52\linewidth]{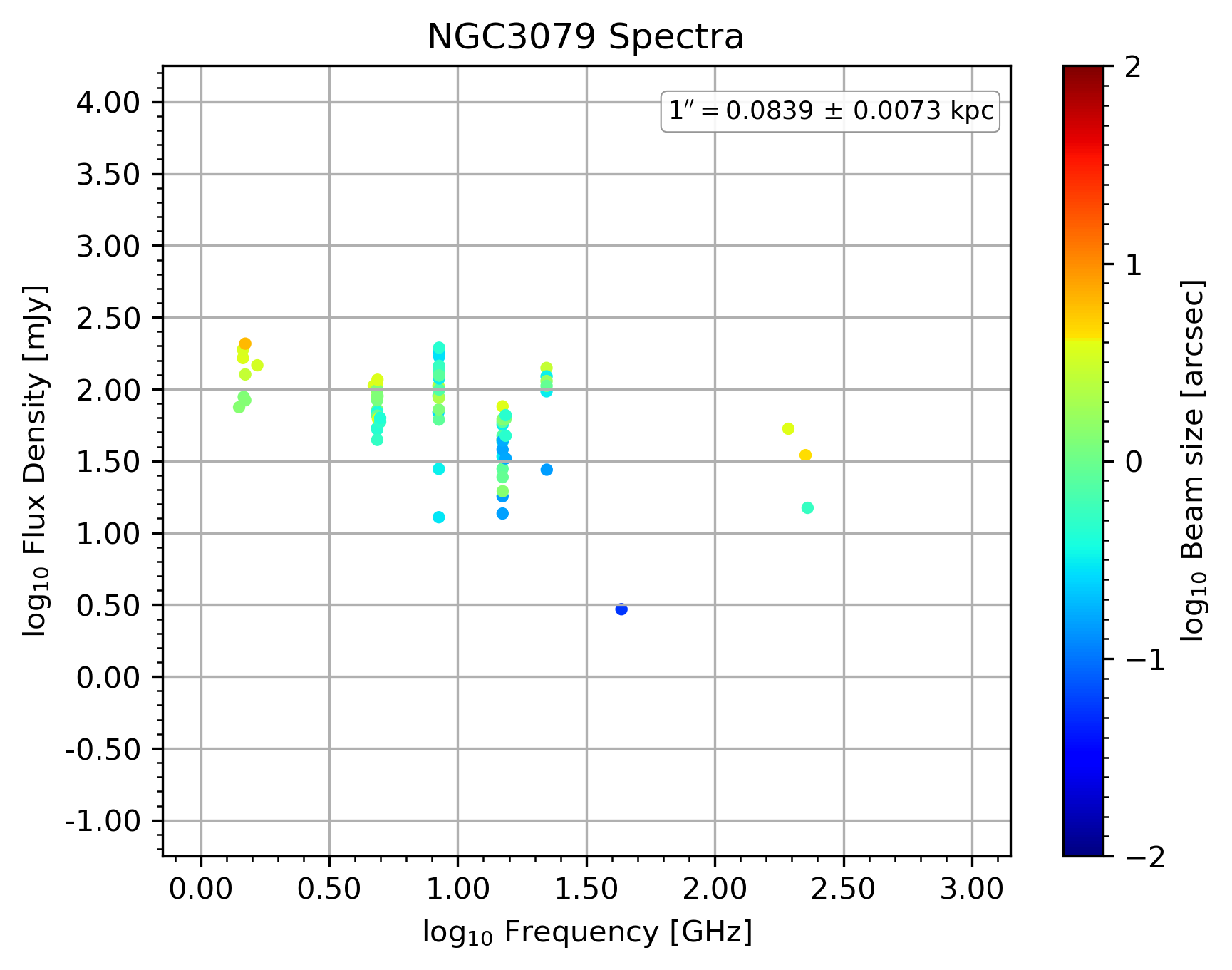}
\caption{Continuum SED of NGC~3079. Color gives the geometric-mean beam FWHM. The inset converts angular to projected physical scale using the adopted distance in Table~\ref{tab:md}.}
\label{fig:sed_3079}
\end{center}
\end{figure*}

\begin{figure*}[t]
\begin{center}
\includegraphics[width=0.52\linewidth]{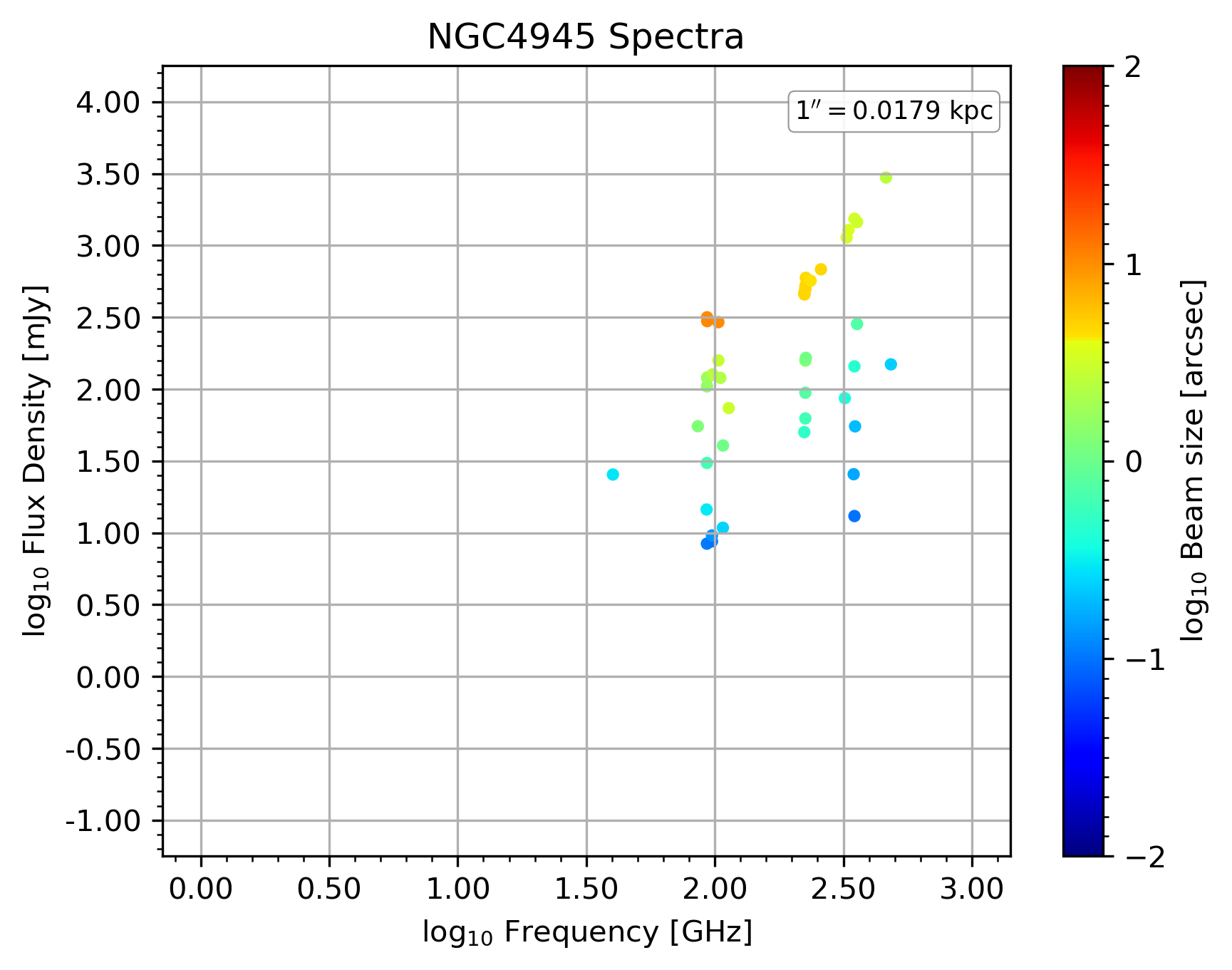}
\caption{Continuum SED of NGC~4945. Color gives the geometric-mean beam FWHM. The inset converts angular to projected physical scale using the adopted distance in Table~\ref{tab:md}.}
\label{fig:sed_4945}
\end{center}
\end{figure*}

The six sources examined in this section have either the largest predicted BHSs or the strongest submm-mm continuum estimates in the sample and therefore set its practical limits. Their continuum spectral energy distributions are shown in Figures~\ref{fig:sed_4258}--\ref{fig:sed_circinus}.

\subsection{NGC~4258}
NGC~4258 is the clearest geometrical BHS candidate in the water maser sample. Its predicted diameter, $0.53860\pm0.00059~\mu$as (statistical) with a separate $\pm0.0054~\mu$as allowance for small imperfections in the Keplerian disk model, comes from the paired VLBI maser-disk fit of \citet{Reid2019} and is the largest in the sample. Its nuclear-region $S_{230}\approx12$~mJy estimate contains the compact core and milliarcsecond jet components (Table~\ref{tab:NGC4258_flux}). An Earth--L2 baseline provides sufficient angular resolution, but shadow-scale correlated flux density, 230~GHz optical thinness, and adequate $(u,v)$ coverage remain to be demonstrated.

Its northern decl. excludes it from ALMA, leaving sparse high-frequency coverage from the SMA, SCUBA-JCMT, and Nobeyama Millimeter Array (NMA) despite extensive VLA archival data.

\begin{figure*}[t]
\begin{center}
\includegraphics[width=0.52\linewidth]{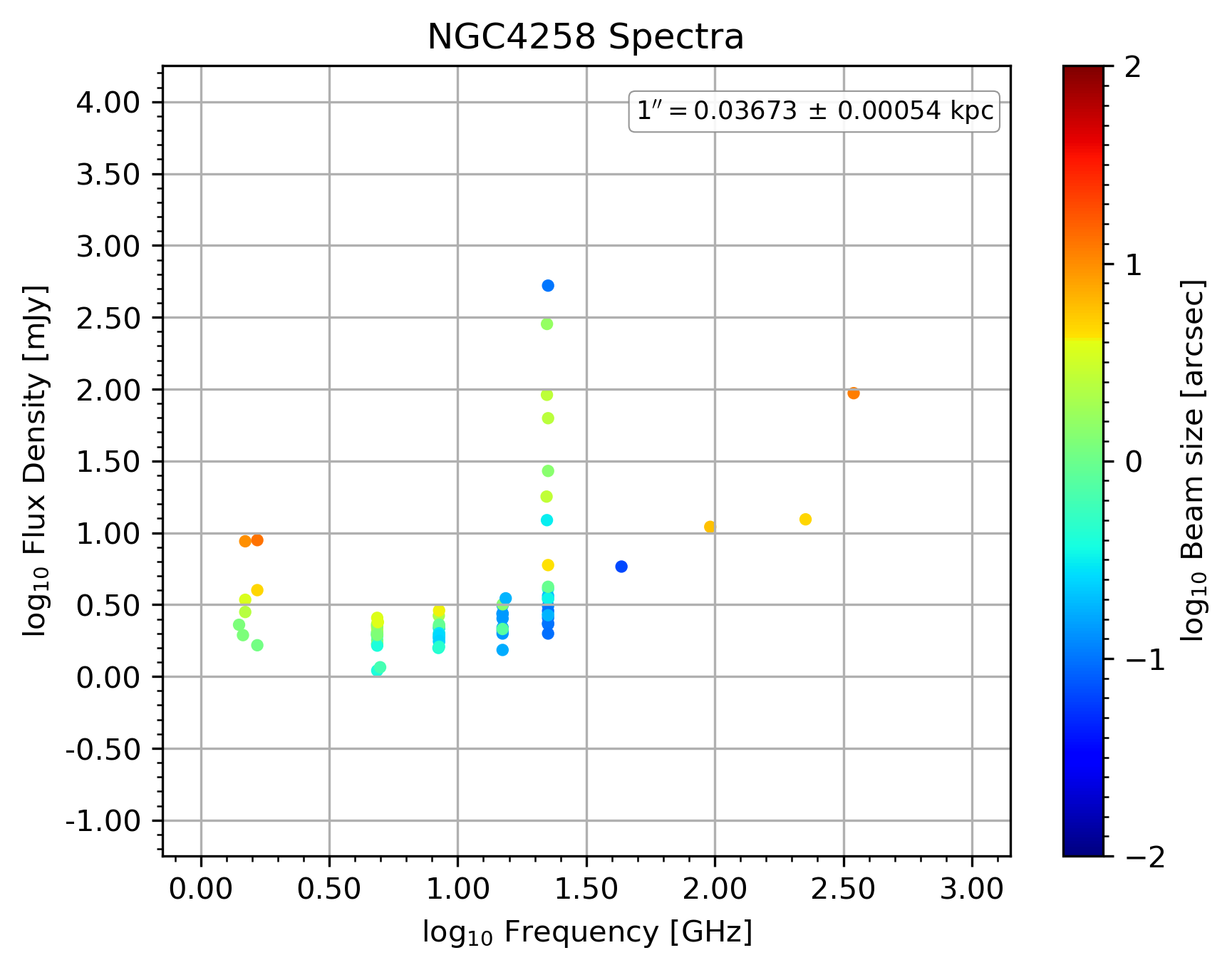}
\caption{Continuum SED of NGC~4258. Color gives the geometric-mean beam FWHM. The inset converts angular to projected physical scale using the adopted distance in Table~\ref{tab:md}.}
\label{fig:sed_4258}
\end{center}
\end{figure*}

Figure~\ref{fig:sed_4258} shows the submm-mm excess: the 96 and 225.5~GHz flux densities exceed the 43.3~GHz measurement. The current sampling does not provide a robust spectral index.

Below 40~GHz, sparse frequency and beam size coverage make the continuum shape difficult to isolate. The measurements nevertheless indicate variability in the regime dominated by extended jet emission, consistent with multi-epoch VLBI studies \citep{Herrnstein1998b}.

Under the tabulated power-law prescription, the modeled maximum thermal-dust contribution is 17\% of $S_{230}$. The 20--2000~K modified-blackbody check gives a more conservative upper limit of 30\% \citep{Doi2013}. The modeled maximum extended-jet contribution is 1.1\% (Table~\ref{tab:NGC4258_flux}). These bounds remain consistent with a substantial AGN-core contribution.

The 22~GHz masers trace a geometrically thin Keplerian disk from $0.110 \pm 0.004$ to $0.29 \pm 0.01$~pc \citep{Gao2016b, Humphreys2008}. The disk is warped, and the red- and blueshifted masers sample different radii and therefore different local inclinations. Table~\ref{tab:md} lists the fitted inclination of $87.05\pm0.09^\circ$ at the 6.1~mas reference radius rather than a single inclination for the entire disk.

For NGC~4258, the $4~\mu$as maser dynamical-center uncertainty is approximately 53 times the maximal spin-dependent offset. The experiment therefore requires substantially improved maser astrometry and registration to the 230~GHz image (Section~\ref{subsec:spin}).

\citet{Lasota1996} used an ADAF model to predict a $\sim$5~mJy flux density at 22~GHz for NGC~4258 \citep[see][]{Herrnstein1998a}. Very Long Baseline Array (VLBA) imaging at 22~GHz detected compact jet components on opposite sides of the maser-disk dynamical center but no continuum emission coincident with the center, yielding a $3\sigma$ upper limit of 0.220~mJy \citep{Herrnstein1998b}. To explain this discrepancy with the ADAF prediction, \citet{Herrnstein2005} proposed two scenarios: (1) the source hosts no ADAF and instead accretes through a geometrically thin disk with low mass supply that extends to the innermost stable circular orbit, or (2) the disk remains geometrically thin down to some transition radius ($< 100~R_{\rm S}$), interior to which it transforms into an ADAF. In the second case, the ADAF synchrotron flux density predicted by \citet{Mahadevan1997} would increase with frequency beyond 22~GHz.

The submm-mm excess found here is consistent with, but does not uniquely establish, the second scenario. Additional evidence compatible with ADAF activity includes twisted nuclear jets at radio, optical, and X-ray wavelengths, potentially produced by magnetically driven bipolar outflows near a quasi-spherical inflow axis. The low X-ray and optical/UV luminosities relative to Eddington are likewise consistent with radiatively inefficient accretion \citep{Narayan1995}.

If the ADAF interpretation is correct, NGC~4258 offers a route to locating the turnover from optically thick to optically thin synchrotron emission, the regime in which photon ring substructure becomes accessible.

\begin{deluxetable*}{lllllll}
\tablecaption{NGC~4258: 230 GHz Continuum Estimates}
\label{tab:NGC4258_flux}
\tablehead{\colhead{Source} & \colhead{Emission} & \colhead{Telescope} & \colhead{$S_{230}$ [mJy]} & \colhead{Input Freq. [GHz]} & \colhead{Beam [\arcsec]} & \colhead{Date}}
\startdata
NGC~4258 & $S_{230}$ & SMA & $12.2 \pm 1.8$ ($12.2$--$12.3$) & 225.51 & 5.63 & 2023/03/16 \\
NGC~4258 & Thermal dust & SCUBA-JCMT & $\leq (2.07 \pm 0.46)$ & 347 & 15 & $\cdots$ \\
NGC~4258 & Extended jet & VLA & $\leq (0.1402 \pm 0.0043)$ & 1.66 & 1.11 & 1997/01/07 \\
\enddata
\tablecomments{The $S_{230}$ value uses $\alpha=-0.7$; parentheses give the range for $-1.0\leq\alpha\leq-0.4$. The dust and jet rows give maximum modeled contributions at 230 GHz, calculated as described in Section~\ref{subsec:contam}.}
\end{deluxetable*}
\subsection{NGC~1194}
NGC~1194 has the second-largest predicted BHS, $0.1253\pm0.0077~\mu$as, based on the VLBI maser-disk $M_\bullet/D$ result of \citet{Kuo2010}. The uncertainty comes from the value compiled by \citet{Gao2016b}, which does not separate statistical from other contributions. Its nearly edge-on disk and large BH mass compensate for its 52.6~Mpc distance, although resolving the BHS would require a baseline 1.4 times Earth--L2. Its nuclear-region continuum estimate is $S_{230}\approx1.5$~mJy (Table~\ref{tab:NGC1194_flux}).

The limiting factor is continuum flux density rather than angular size: at fixed spectral luminosity, a source at 52.6~Mpc would have about 1/48 of NGC~4258's flux density. The modeled maximum thermal-dust and extended-jet contributions are 21\% and 3.7\% of $S_{230}$, respectively (Table~\ref{tab:NGC1194_flux}). The spectrum remains broadly flat from $\sim$5--310~GHz, a common signature of compact AGN-core emission \citep{Blandford1979,Shabala2012}.

Its 22~GHz masers trace a Keplerian disk from 0.54$^{+0.25}_{-0.03}$ to 1.33$^{+0.06}_{-0.97}$~pc \citep{Kuo2010}. General-relativistic modeling, also applied to NGC~4258 \citep{Nucamendi2021}, predicts a gravitational-redshift contribution of about 2.3~km~s$^{-1}$ for the innermost observed maser feature in NGC~1194 \citep{GonzalezJuarez2024}. NGC~1194 also hosts 183~GHz water masers, but these lines are far fainter than the $\sim$1~Jy 22~GHz emission and are concentrated near the systemic features \citep{Pesce2023}.

\begin{deluxetable*}{lllllll}
\tablecaption{NGC~1194: 230 GHz Continuum Estimates}
\label{tab:NGC1194_flux}
\tablehead{\colhead{Source} & \colhead{Emission} & \colhead{Telescope} & \colhead{$S_{230}$ [mJy]} & \colhead{Input Freq. [GHz]} & \colhead{Beam [\arcsec]} & \colhead{Date}}
\startdata
NGC~1194 & $S_{230}$ & ALMA & $1.516 \pm 0.053$ ($1.39$--$1.66$) & 310.04 & 0.0228 & 2025/08/09 \\
NGC~1194 & Thermal dust & ALMA & $\leq (0.321 \pm 0.011)$ & 310.04 & 0.0228 & 2025/08/09 \\
NGC~1194 & Extended jet & VLA & $\leq (0.0563 \pm 0.0077)$ & 1.49 & 14.9 & 1988/05/03 \\
\enddata
\tablecomments{The $S_{230}$ value uses $\alpha=-0.7$; parentheses give the range for $-1.0\leq\alpha\leq-0.4$. The dust and jet rows give maximum modeled contributions at 230 GHz, calculated as described in Section~\ref{subsec:contam}.}
\end{deluxetable*}

\subsection{NGC~1068}
NGC~1068 combines a complex disk--outflow system with the third-largest predicted BHS in the sample. The inclination-corrected mass estimate of \citet{Gallimore2024}, placed on the Cepheid distance of \citet{Markham2026}, gives a predicted diameter of $0.11825~\mu\text{as}$. The published fit term corresponds to $\pm0.00071~\mu\text{as}$, while the complete model uncertainty remains to be quantified. The maser disk is nearly edge-on, and its polarization constrains the magnetic-field geometry \citep{Gallimore2024}. Resolving the BHS would require a baseline about 1.5 times Earth--L2; its 230~GHz continuum estimate is $S_{230}\approx10.7$~mJy (Table~\ref{tab:NGC1068_flux}).

\begin{figure*}[t]
\begin{center}
\includegraphics[width=0.52\linewidth]{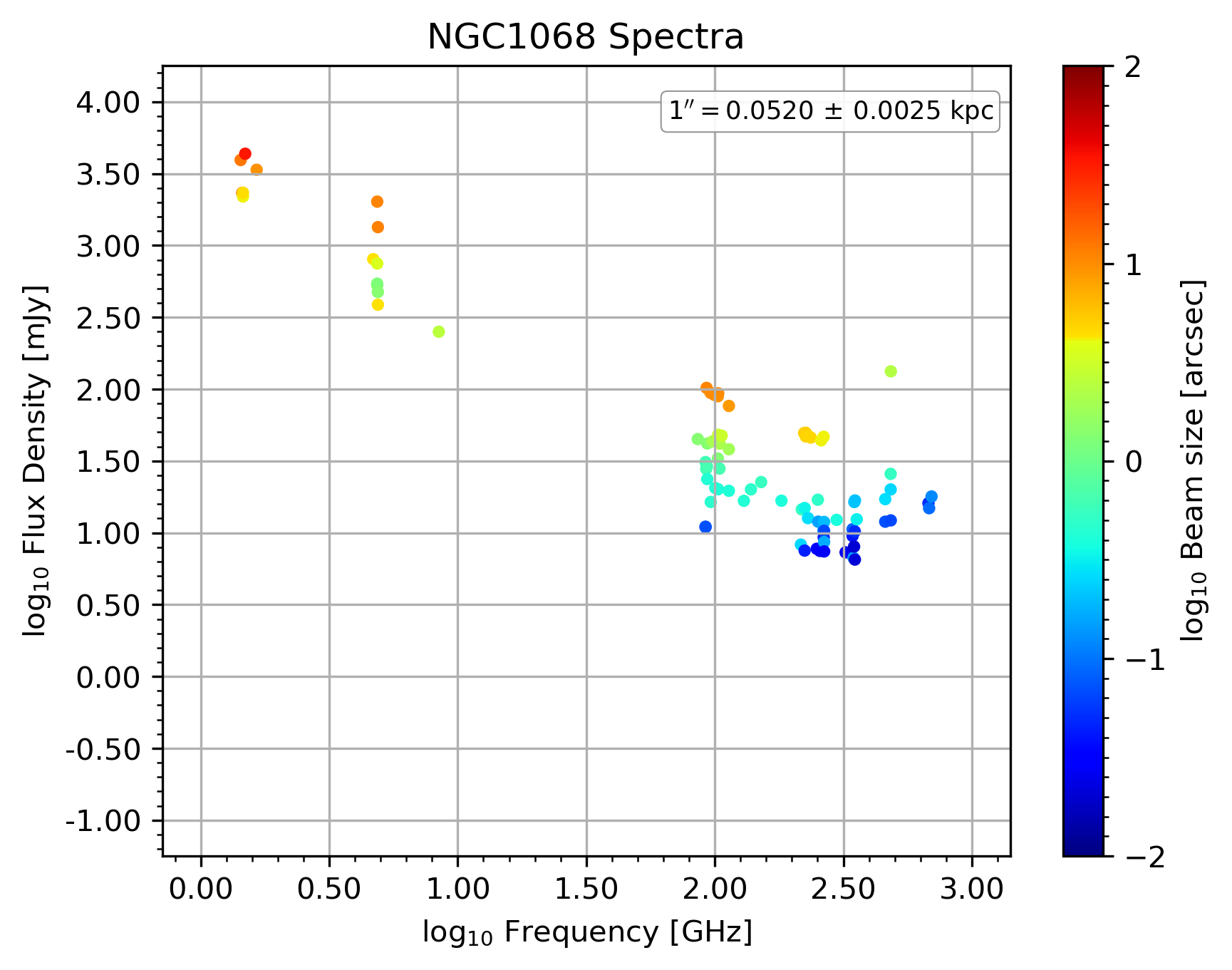}
\caption{Continuum SED of NGC~1068. Color gives the geometric-mean beam FWHM. The inset converts angular to projected physical scale using the adopted distance in Table~\ref{tab:md}.}
\label{fig:sed_1068}
\end{center}
\end{figure*}

The compiled data decline from centimeter wavelengths before showing a 200--700~GHz excess (Figure~\ref{fig:sed_1068}). \citet{Mutie2025} attribute this excess to AGN-core synchrotron emission that is optically thick from 200--500~GHz and increasingly transparent toward 700~GHz. The modeled maximum extended-jet contribution is about 20\% of $S_{230}$ \citep{Gallimore2024}, while the modeled maximum thermal-dust contribution is 0.0013\% (Table~\ref{tab:NGC1068_flux}).

The 22~GHz masers trace a Keplerian disk at 0.35--1~pc \citep{Gallimore2023}, while searches at 183 and 321~GHz have yielded no detections \citep{Pesce2023,Pesce2016}. The maser and molecular disks are rotated by about 30\textdegree{} relative to the elongated plasma disk, and maser filaments displaced above the fitted disk plane may trace material lifted toward the base of the outflow \citep{Gallimore2024}.

Two additional 22~GHz maser populations lie outside the main orbiting disk. One forms an annular structure consistent with a jet colliding with a molecular cloud \citep{Morishima2022, Gallimore2023}; the other traces a wide-angle outflow similar to that in Circinus. Its maser substructure and disk--outflow interface make NGC~1068 a promising laboratory for dynamical models aimed at improving the maser dynamical-center estimate. Polarization constraints also offer a future test of whether MAD or SANE models can reproduce its magnetic geometry and disk--outflow alignment \citep{Gallimore2024,Ressler2023}.

\begin{deluxetable*}{lllllll}
\tablecaption{NGC~1068: 230 GHz Continuum Estimates}
\label{tab:NGC1068_flux}
\tablehead{\colhead{Source} & \colhead{Emission} & \colhead{Telescope} & \colhead{$S_{230}$ [mJy]} & \colhead{Input Freq. [GHz]} & \colhead{Beam [\arcsec]} & \colhead{Date}}
\startdata
NGC~1068 & $S_{230}$ & ALMA & $10.724 \pm 0.048$ ($9.47$--$12.1$) & 348.32 & 0.0139 & 2023/07/28 \\
NGC~1068 & Thermal dust & ALMA & $\leq (0.0001377 \pm 0.0000081)$ & 484.25 & 2.56 & 2018/05/13 \\
NGC~1068 & Extended jet & HSA & $\leq (2.16 \pm 0.31)$ & 21.9 & 0.000888 & 2021/11/14 \\
\enddata
\tablecomments{The $S_{230}$ value uses $\alpha=-0.7$; parentheses give the range for $-1.0\leq\alpha\leq-0.4$. The dust and jet rows give maximum modeled contributions at 230 GHz, calculated as described in Section~\ref{subsec:contam}.}
\end{deluxetable*}

\subsection{NGC~3079}
NGC~3079 combines a strong nuclear-region continuum estimate, $S_{230}\approx14.9$~mJy, with the smallest nominal BHS diameter, $0.0119~\mu$as. The maser kinematics indicate approximately $2\times10^6~M_\odot$ within 0.4~pc, which sets the nominal BHS diameter. Its flat rotation curve and thick, self-gravitating disk imply that some of this mass may reside in the disk rather than the SMBH, so the nominal BHS diameter could be too large; this contribution is not quantified. In the opposite direction, the conditional turbulent-support model of \citet{Kuo2020} would increase the inferred SMBH mass and BHS diameter by 34\%. Table~\ref{tab:md} keeps this one-sided model shift separate. Matching the nominal BHS diameter would require a baseline about 15.1 times Earth--L2.

The modeled maximum extended-jet contribution is 6.1\% of $S_{230}$ \citep{Sawada2000}. No thermal-dust estimate is available because there is no suitable higher-frequency continuum measurement (Table~\ref{tab:NGC3079_flux}).

VLBA maps place the 22~GHz masers at radii of roughly 0.4--1.3~pc \citep{Yamauchi2004,Kondratko2005}. RadioAstron has also detected 22~GHz H$_2$O megamaser emission from NGC~3079 on Earth--space baselines. Unlike NGC~4258, its features weaken or disappear on longer baselines, implying that much of the emission is spatially extended and resolved out. \citet{Baan2017} interpret the detected blueshifted masers as shocked gas in the nuclear interstellar medium, possibly linked to nuclear outflow or starburst activity, rather than compact clumps in a clean Keplerian disk.

NGC~3079 shows nuclear 183~GHz emission whose strongest features fall within velocities occupied by its 22~GHz masers, consistent with related disk or outflow structures. \citet{Humphreys2005} also reported tentative 439~GHz emission peaking within this velocity range. Available spatial resolution cannot establish whether it arises in disk or outflow gas. Confirming and mapping this transition would provide an additional probe of excitation conditions in nuclear molecular gas.

The maser map further departs from a geometrically thin disk picture: components show structure orthogonal to the main elongation and large velocity dispersion in compact sky regions. \citet{Kondratko2005} interpret the system as a disk interacting with a jet misaligned from the disk rotation axis. The morphology resembles NGC~1068, where water masers have been modeled as spiral-arm structures lifted away from the disk midline by a disk wind \citep{Gallimore2024}.

\begin{deluxetable*}{lllllll}
\tablecaption{NGC~3079: 230 GHz Continuum Estimates}
\label{tab:NGC3079_flux}
\tablehead{\colhead{Source} & \colhead{Emission} & \colhead{Telescope} & \colhead{$S_{230}$ [mJy]} & \colhead{Input Freq. [GHz]} & \colhead{Beam [\arcsec]} & \colhead{Date}}
\startdata
NGC~3079 & $S_{230}$ & NOEMA & $14.9$ ($14.9$--$14.9$) & 229.66 & 0.503 & 2018/02/07 \\
NGC~3079 & Extended jet & VLBA+VLA & $\leq (0.91 \pm 0.19)$ & 1.4 & 0.0137 & 1996/10/20 \\
\enddata
\tablecomments{The $S_{230}$ value uses $\alpha=-0.7$; parentheses give the range for $-1.0\leq\alpha\leq-0.4$. The jet row gives the maximum modeled contribution at 230 GHz, calculated as described in Section~\ref{subsec:contam}. The adopted 229.66 GHz NOEMA measurement is reported without a flux-density uncertainty.}
\end{deluxetable*}

\subsection{NGC~4945}
\label{sec:NGC4945}
From a sparse VLBI maser map, \citet{Greenhill1997} inferred nearly edge-on circular rotation and an approximate mass enclosed within $\sim0.3$~pc. Treating this enclosed mass as the SMBH mass gives the nominal BHS diameter of $0.0388~\mu$as reported here. The asymmetric maser distribution makes this interpretation uncertain. Separately, the conditional turbulent-support model of \citet[Appendix~A]{Kuo2020} would increase the inferred mass and BHS diameter by 24\%; Table~\ref{tab:md} keeps this one-sided model shift separate. No fitted inclination was reported. Matching the nominal BHS diameter would require a baseline about 4.6 times Earth--L2. Its nuclear-region continuum estimate is $S_{230}\approx17.5$~mJy (Table~\ref{tab:NGC4945_flux}).

NGC~4945 is a deeply buried AGN embedded in an active starburst. Its hard X-ray emission provides the clearest AGN diagnostic: the source is among the brightest X-ray sources on the sky and varies on $\sim$1 day timescales \citep{Emig2020, Hagiwara2021, Perez-Beaupuits2011, Marconi2000}. At centimeter wavelengths, however, the continuum is dominated by star formation and free-free radiation \citep{Emig2020, Bendo2016}, consistent with the high visual extinction of the host \citep{Moorwood1996, Goulding2009}. Supernova feedback and feedback from stellar winds may help keep the obscuring gas from collapsing into a geometrically thin disk \citep{Fabian1998, Marconi2000}, and mid-infrared spectra reveal an extended, fragmented, star-forming molecular ring or disk around the buried AGN \citep{Spoon2003}.

The modeled maximum thermal-dust contribution is 0.45\% of $S_{230}$. No extended-jet estimate is adopted because the centimeter continuum is dominated by host-galaxy star formation and free--free radiation (Table~\ref{tab:NGC4945_flux}).

At 22~GHz, the masers span $\sim$0.7~pc; the galaxy's low declination restricted the VLBI hour-angle coverage and increased the positional uncertainties \citep{Greenhill1997}.

Higher-frequency water masers are also present at 183 and 321~GHz \citep{Humphreys2016, Hagiwara2016, Hagiwara2021, Pesce2016}. The 321~GHz peaks do not map one-to-one onto the 22~GHz velocities; instead, they appear where the 22~GHz emission drops off. Because 321~GHz masers prefer higher kinetic temperatures, they may arise at smaller disk radii than the 22~GHz transition. Across the two simultaneous ALMA line-and-continuum epochs analyzed by \citet{Hagiwara2021}, the 321~GHz maser and continuum both weakened, suggesting---but not establishing---a connection between their variability.

\begin{deluxetable*}{lllllll}
\tablecaption{NGC~4945: 230 GHz Continuum Estimates}
\label{tab:NGC4945_flux}
\tablehead{\colhead{Source} & \colhead{Emission} & \colhead{Telescope} & \colhead{$S_{230}$ [mJy]} & \colhead{Input Freq. [GHz]} & \colhead{Beam [\arcsec]} & \colhead{Date}}
\startdata
NGC~4945 & $S_{230}$ & ALMA & $17.5 \pm 2.5$ ($15.5$--$19.9$) & 349.4 & 0.0783 & 2023/05/19 \\
NGC~4945 & Thermal dust & ALMA & $\leq (0.0782 \pm 0.0032)$ & 258.8 & 5.61 & 2018/09/06 \\
\enddata
\tablecomments{The $S_{230}$ value uses $\alpha=-0.7$; parentheses give the range for $-1.0\leq\alpha\leq-0.4$. The dust row gives the maximum modeled contribution at 230 GHz, calculated as described in Section~\ref{subsec:contam}. No extended-jet row is included because the centimeter continuum is dominated by host-galaxy star formation and free--free radiation rather than an extended AGN jet.}
\end{deluxetable*}

\subsection{Circinus}
At 22~GHz, VLBI imaging of Circinus shows a warped, nearly Keplerian disk and a wide-angle outflow \citep{Greenhill2003,McCallum2009}. The disk kinematics give the nominal central-mass-to-distance ratio used to predict a BHS diameter of $0.04152~\mu$as. \citet{Greenhill2003} estimated that the disk mass is probably less than 24\% of the central mass. A non-negligible disk contribution would make the nominal SMBH mass and BHS diameter too large, but this contribution was not measured separately and is therefore not assigned a numerical correction. In the opposite direction, the conditional turbulent-support model of \citet{Kuo2020} would increase the BHS diameter by 3\%; Table~\ref{tab:md} lists this one-sided model shift separately. Other contributions to the uncertainty remain incompletely quantified. The disk extends from $0.11\pm0.02$ to $\sim0.40$~pc and the outflow to $\sim1$~pc \citep{Greenhill2003,Hagiwara2021}. Matching the nominal BHS diameter would require a baseline about 4.3 times Earth--L2; its nuclear-region continuum estimate is $S_{230}\approx15.5$~mJy (Table~\ref{tab:circinus_flux}).

Outflow masers probably originate within 0.1~pc of the central engine. The warped disk appears to limit the directions in which the outflow escapes and thereby shape the larger-scale ionization cone \citep{Greenhill2003}.

\begin{figure*}[t]
\begin{center}
\includegraphics[width=0.52\linewidth]{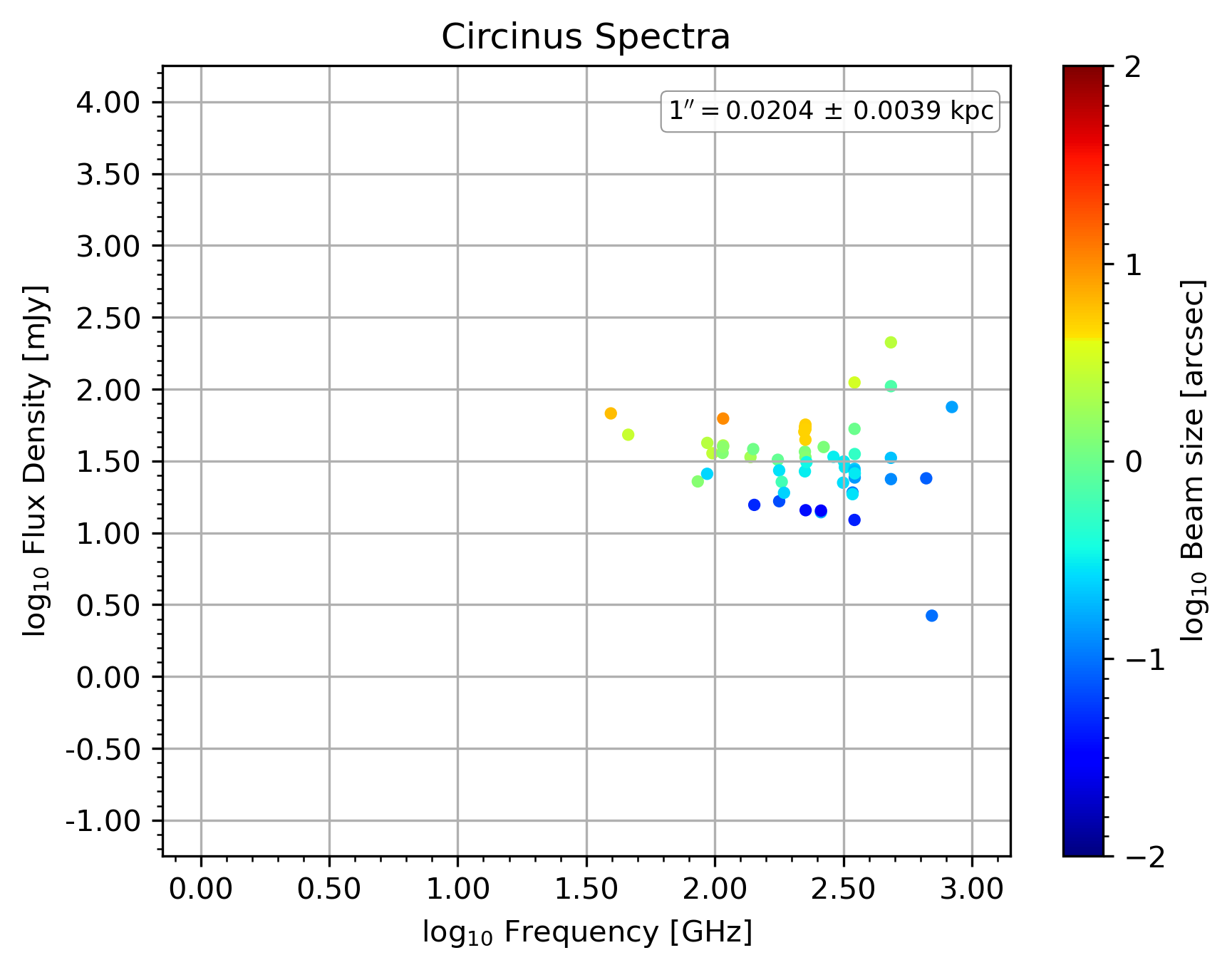}
\caption{Continuum SED of Circinus. Color gives the geometric-mean beam FWHM. The inset converts angular to projected physical scale using the adopted distance in Table~\ref{tab:md}.}
\label{fig:sed_circinus}
\end{center}
\end{figure*}

Circinus lies outside the accessible sky of the SMA and VLA. Its low Galactic latitude causes high extinction \citep{Schlafly2011}, while diffractive interstellar scintillation in the Milky Way produces rapid variability in the maser features \citep{McCallum2009}. Figure~\ref{fig:sed_circinus} suggests different frequency trends at different angular resolutions: larger-beam measurements generally rise toward higher frequencies, while the smallest-beam measurements decline between approximately 180 and 350~GHz. The modeled maximum thermal-dust contribution is 0.0027\% of $S_{230}$. The modeled maximum extended-jet contribution is about 46\% of $S_{230}$ \citep{Elmouttie1998} (Table~\ref{tab:circinus_flux}). These estimates are consistent with a substantial AGN-core contribution to the submm-mm continuum.

Circinus also hosts 183 and 321~GHz water masers, which probe different excitation conditions. The strong 183~GHz emission is distributed similarly to the 22~GHz masers and may trace the disk at larger radii \citep{Pesce2023}. The 321~GHz features vary strongly and may trace the circumnuclear disk or outflow \citep{Hagiwara2013,Pesce2016,Hagiwara2021}.

\begin{deluxetable*}{lllllll}
\tablecaption{Circinus: 230 GHz Continuum Estimates}
\label{tab:circinus_flux}
\tablehead{\colhead{Source} & \colhead{Emission} & \colhead{Telescope} & \colhead{$S_{230}$ [mJy]} & \colhead{Input Freq. [GHz]} & \colhead{Beam [\arcsec]} & \colhead{Date}}
\startdata
Circinus & $S_{230}$ & ALMA & $15.500 \pm 0.089$ ($15.0$--$16.1$) & 258.83 & 0.0206 & 2019/06/06 \\
Circinus & Thermal dust & ALMA & $\leq (0.000411 \pm 0.000018)$ & 484.47 & 2.77 & 2018/10/27 \\
Circinus & Extended jet & ATCA & $\leq (7.11 \pm 0.71)$ & 2.37 & 3.25 & 1994/03-1996/12 \\
\enddata
\tablecomments{The $S_{230}$ value uses $\alpha=-0.7$; parentheses give the range for $-1.0\leq\alpha\leq-0.4$. The dust and jet rows give maximum modeled contributions at 230 GHz, calculated as described in Section~\ref{subsec:contam}.}
\end{deluxetable*}

\clearpage

\section{Future Work}
\label{sec:future}

Although 22~GHz water masers will remain the most commonly observed transition due to their relatively broad range of excitation densities and temperatures, water masers at transitions beyond 22~GHz provide a valuable diagnostic of the physical conditions and excitation mechanisms within AGN accretion disks, and continued searches for such lines remain essential. In addition to the high-frequency detections and non-detections summarized in Section~\ref{subsec:spin}, \citet{Pesce2023} report 183~GHz detections in ESO~558--G009, IC~2560, UGC~6093, CGCG~074-064, and NGC~5765b, which are sources in the 21-source sample not discussed individually here. If sufficiently compact higher-frequency water masers can be mapped with VLBI and registered to continuum in the same observing band, they could remove one cross-band calibration obstacle for the BHS--maser registration problem, but they would not by themselves solve the spin-offset measurement: the maser dynamical center would still need to be located with sub-$\mu$as precision and tied to the continuum image.

Another component of the relative-astrometry problem is frequency-dependent core shift in a calibrator common to both bands. This effect may introduce a systematic offset between the 22~GHz and 230~GHz reference frames; its magnitude is not quantified here and therefore remains an unmodeled source of systematic uncertainty.

More sophisticated dynamical models of maser disks and disk--outflow interfaces may improve dynamical-center estimates. Magnetized-flow models spanning MAD and SANE states could separately test how closely maser-disk and inner accretion-flow orientations are aligned.

Future reverberation mapping of megamaser AGN with observable broad lines could provide distance-independent mass estimates that, combined with maser measurements of $M_\bullet/D$, could yield independent distance estimates.

\section{Conclusion}
\label{sec:conclusion}
Within our 21-source megamaser sample, NGC~4258 remains the leading geometrical and arcsecond-scale flux-density candidate for BHS imaging: its predicted BHS is approximately $0.54~\mu\mathrm{as}$, and an Earth--L2 baseline delivers $\theta\approx0.18~\mu\mathrm{as}$. Its $S_{230}\approx12$~mJy estimate does not establish shadow-scale correlated flux density. A sufficiently sensitive northern ground array and 230~GHz optical thinness must be demonstrated before robust detectability or imaging feasibility can be claimed. The compiled continuum measurements show a submm-mm excess in NGC~4258. Its spectral shape is consistent with a scenario in which the disk remains geometrically thin down to a transition radius of $\lesssim 100\,R_S$ and becomes an ADAF interior to that radius, although the current sampling does not uniquely determine the accretion structure \citep{Lasota1996, Herrnstein1998a, Herrnstein2005}.

A central contribution of this work is a spin observable: the offset between the BHS and the megamaser disk dynamical center. The maximal offset is $\theta_{\mathrm{shift}} \simeq 0.14\theta_{\mathrm{BHS}}$; for NGC~4258, this implies $\theta_{\mathrm{shift}} \approx 0.075~\mu\mathrm{as}$. At 230~GHz, an Earth--Moon baseline provides the required $1\sigma$ BHS centroid precision with a SNR of $\sim 4.1$, while a robust $3\sigma$ detection would require SNR$\sim 12$. This formal BHS centroid precision does not make the spin-offset experiment feasible. The $4~\mu\mathrm{as}$ astrometric uncertainty derived from the 22~GHz water-maser coordinates is approximately 53 times the maximal offset. The dynamical-center precision must reach $\lesssim 0.075~\mu\mathrm{as}$, and the center must also be registered to the 230~GHz BHS image before additional cross-band systematics such as calibrator core shift are considered. Longer SVLBI baselines therefore do not remove the dominant blocker; a plausible path forward is improved modeling of maser substructure and disk/outflow physics, with NGC~1068 providing a particularly promising laboratory for that work.

All other sample sources with adopted primary BHS sizes require baselines extending beyond Earth--L2. Baselines reaching Earth--L4/L5 recover all adopted primary BHS-scale fringe spacings for the water maser sample, but angular resolution alone does not establish imaging feasibility. Five sources in the sample have $S_{230}\gtrsim10$~mJy.

Beyond NGC~4258, NGC~1194 and NGC~1068 have the next-largest adopted BHS sizes, while NGC~3079, NGC~4945, and Circinus provide the strongest continuum estimates among the remaining detailed cases. Their individual discussions show how, where relevant, gravitational redshift, magnetic geometry, disk warping and outflows, and star formation shape the interpretation of these targets.

\section*{Acknowledgments}
We thank the anonymous referees for the careful and constructive comments. This work was supported by the National Science Foundation (NSF) under grant AST-2034306. We thank D.~W.~Pesce for supplying paired mass--distance ratio summaries from the published chains and uncertainty guidance. OpenAI's Codex was used for manuscript and software preparation. This research made use of the NASA/IPAC Extragalactic Database (NED; \citealt{NED2019}) and of observations obtained with the Submillimeter Array (SMA) under programs 2022B-H002 and 2021B-H004, as well as Atacama Large Millimeter/submillimeter Array (ALMA) and Karl G. Jansky Very Large Array (VLA) data retrieved from their respective science archives. We acknowledge that the SMA is located on Maunakea, a site of profound cultural and spiritual significance to the Kānaka Maoli; that ALMA is located on the ancestral lands of the Likan Antai (Atacameño) People; and that the VLA is located on land of longstanding cultural significance to Indigenous peoples of the Socorro region. We are grateful for the privilege of studying the cosmos from these lands. The SMA is a joint project between the Smithsonian Astrophysical Observatory and the Academia Sinica Institute of Astronomy and Astrophysics and is funded by the Smithsonian Institution and the Academia Sinica. The National Radio Astronomy Observatory (NRAO) and Green Bank Observatory are facilities of the U.S. National Science Foundation operated under cooperative agreement by Associated Universities, Inc. This paper makes use of the following ALMA data: \nolinkurl{ADS/JAO.ALMA#2012.1.00139.S}, \nolinkurl{ADS/JAO.ALMA#2012.1.00332.S}, \nolinkurl{ADS/JAO.ALMA#2012.1.00479.S}, \nolinkurl{ADS/JAO.ALMA#2012.1.00912.S}, \nolinkurl{ADS/JAO.ALMA#2013.1.00014.S}, \nolinkurl{ADS/JAO.ALMA#2013.1.00055.S}, \nolinkurl{ADS/JAO.ALMA#2013.1.00060.S}, \nolinkurl{ADS/JAO.ALMA#2013.1.00099.S}, \nolinkurl{ADS/JAO.ALMA#2013.1.00111.S}, \nolinkurl{ADS/JAO.ALMA#2013.1.00188.S}, \nolinkurl{ADS/JAO.ALMA#2013.1.00221.S}, \nolinkurl{ADS/JAO.ALMA#2013.1.00247.S}, \nolinkurl{ADS/JAO.ALMA#2013.1.00279.S}, \nolinkurl{ADS/JAO.ALMA#2013.1.00824.S}, \nolinkurl{ADS/JAO.ALMA#2013.1.01151.S}, \nolinkurl{ADS/JAO.ALMA#2013.1.01307.S}, \nolinkurl{ADS/JAO.ALMA#2015.1.00086.S}, \nolinkurl{ADS/JAO.ALMA#2015.1.00295.S}, \nolinkurl{ADS/JAO.ALMA#2015.1.00455.S}, \nolinkurl{ADS/JAO.ALMA#2015.1.00476.S}, \nolinkurl{ADS/JAO.ALMA#2015.1.00960.S}, \nolinkurl{ADS/JAO.ALMA#2015.1.01144.S}, \nolinkurl{ADS/JAO.ALMA#2015.1.01286.S}, \nolinkurl{ADS/JAO.ALMA#2016.1.00023.S}, \nolinkurl{ADS/JAO.ALMA#2016.1.00052.S}, \nolinkurl{ADS/JAO.ALMA#2016.1.00150.S}, \nolinkurl{ADS/JAO.ALMA#2016.1.00176.S}, \nolinkurl{ADS/JAO.ALMA#2016.1.00232.S}, \nolinkurl{ADS/JAO.ALMA#2016.1.01135.S}, \nolinkurl{ADS/JAO.ALMA#2016.1.01279.S}, \nolinkurl{ADS/JAO.ALMA#2016.1.01553.S}, \nolinkurl{ADS/JAO.ALMA#2016.1.01613.S}, \nolinkurl{ADS/JAO.ALMA#2016.2.00098.S}, \nolinkurl{ADS/JAO.ALMA#2017.1.00002.S}, \nolinkurl{ADS/JAO.ALMA#2017.1.00082.S}, \nolinkurl{ADS/JAO.ALMA#2017.1.00236.S}, \nolinkurl{ADS/JAO.ALMA#2017.1.00575.S}, \nolinkurl{ADS/JAO.ALMA#2017.1.00586.S}, \nolinkurl{ADS/JAO.ALMA#2017.1.01226.S}, \nolinkurl{ADS/JAO.ALMA#2017.1.01666.S}, \nolinkurl{ADS/JAO.ALMA#2018.1.00007.S}, \nolinkurl{ADS/JAO.ALMA#2018.1.00037.S}, \nolinkurl{ADS/JAO.ALMA#2018.1.00064.S}, \nolinkurl{ADS/JAO.ALMA#2018.1.00321.S}, \nolinkurl{ADS/JAO.ALMA#2018.1.00581.S}, \nolinkurl{ADS/JAO.ALMA#2018.1.00986.S}, \nolinkurl{ADS/JAO.ALMA#2018.1.01135.S}, \nolinkurl{ADS/JAO.ALMA#2018.1.01236.S}, \nolinkurl{ADS/JAO.ALMA#2018.1.01321.S}, \nolinkurl{ADS/JAO.ALMA#2018.1.01506.S}, \nolinkurl{ADS/JAO.ALMA#2018.1.01684.S}, \nolinkurl{ADS/JAO.ALMA#2018.A.00062.S}, \nolinkurl{ADS/JAO.ALMA#2019.1.00013.S}, \nolinkurl{ADS/JAO.ALMA#2019.1.00014.S}, \nolinkurl{ADS/JAO.ALMA#2019.1.00026.S}, \nolinkurl{ADS/JAO.ALMA#2019.1.00130.S}, \nolinkurl{ADS/JAO.ALMA#2019.1.00416.S}, \nolinkurl{ADS/JAO.ALMA#2019.1.00763.L}, \nolinkurl{ADS/JAO.ALMA#2019.1.00854.S}, \nolinkurl{ADS/JAO.ALMA#2019.1.01230.S}, \nolinkurl{ADS/JAO.ALMA#2019.1.01303.S}, \nolinkurl{ADS/JAO.ALMA#2019.1.01540.S}, \nolinkurl{ADS/JAO.ALMA#2019.2.00129.S}, \nolinkurl{ADS/JAO.ALMA#2021.1.00026.S}, \nolinkurl{ADS/JAO.ALMA#2021.1.00279.S}, \nolinkurl{ADS/JAO.ALMA#2021.1.00783.S}, \nolinkurl{ADS/JAO.ALMA#2021.1.00812.S}, \nolinkurl{ADS/JAO.ALMA#2021.1.01150.S}, \nolinkurl{ADS/JAO.ALMA#2021.1.01183.S}, \nolinkurl{ADS/JAO.ALMA#2021.1.01295.S}, \nolinkurl{ADS/JAO.ALMA#2021.1.01644.S}, \nolinkurl{ADS/JAO.ALMA#2021.2.00049.S}, \nolinkurl{ADS/JAO.ALMA#2021.2.00077.S}, \nolinkurl{ADS/JAO.ALMA#2022.1.00005.S}, \nolinkurl{ADS/JAO.ALMA#2022.1.00017.S}, \nolinkurl{ADS/JAO.ALMA#2022.1.00190.S}, \nolinkurl{ADS/JAO.ALMA#2022.1.00222.S}, \nolinkurl{ADS/JAO.ALMA#2022.1.00343.S}, \nolinkurl{ADS/JAO.ALMA#2022.1.01002.S}, \nolinkurl{ADS/JAO.ALMA#2022.1.01255.S}, \nolinkurl{ADS/JAO.ALMA#2022.1.01576.S}, \nolinkurl{ADS/JAO.ALMA#2023.1.00026.S}, \nolinkurl{ADS/JAO.ALMA#2023.1.00064.S}, \nolinkurl{ADS/JAO.ALMA#2023.1.01517.S}, \nolinkurl{ADS/JAO.ALMA#2023.1.01660.S}, \nolinkurl{ADS/JAO.ALMA#2024.1.00006.S}, \nolinkurl{ADS/JAO.ALMA#2024.1.00211.S}, \nolinkurl{ADS/JAO.ALMA#2024.1.00480.S}, \nolinkurl{ADS/JAO.ALMA#2024.1.00821.S}, \nolinkurl{ADS/JAO.ALMA#2024.1.00968.S}, \nolinkurl{ADS/JAO.ALMA#2024.1.01245.S}, \nolinkurl{ADS/JAO.ALMA#2024.1.01508.S}. ALMA is a partnership of ESO (representing its member states), NSF (USA), and NINS (Japan), together with NRC (Canada), NSTC and ASIAA (Taiwan), and KASI (Republic of Korea), in cooperation with the Republic of Chile. The Joint ALMA Observatory is operated by ESO, AUI/NRAO, and NAOJ.

\facilities{ALMA, ATCA, GBT, IRAM:NOEMA, JCMT (SCUBA), NoMA, SMA, VLA, VLBA}
\software{WM-AGN Analysis Code (\citealt{Burridge2026})}

\end{document}